\RequirePackage[switch, columnwise, running, mathlines, displaymath, mathlines]{lineno}
\documentclass[twocolumn,nofootinbib,superscriptaddress]{revtex4-2}
\usepackage[utf8]{inputenc}
\usepackage{amsmath}
\usepackage{gensymb}
\usepackage{amssymb}
\usepackage{graphicx}
\usepackage[dvipsnames]{xcolor}
\usepackage{hyperref}
\usepackage[normalem]{ulem}
\usepackage{aas_macros}
\usepackage{bm}
\usepackage{savesym}
\savesymbol{tablenum}
\usepackage{siunitx}
\restoresymbol{SIX}{tablenum}
\usepackage{enumitem}

\usepackage{tabularx}
\usepackage{colortbl}
\renewcommand{\arraystretch}{1.2}

\makeatletter
    \def\CT@@do@color{%
      \global\let\CT@do@color\relax
            \@tempdima\wd\z@
            \advance\@tempdima\@tempdimb
            \advance\@tempdima\@tempdimc
    \advance\@tempdimb\tabcolsep
    \advance\@tempdimc\tabcolsep
    \advance\@tempdima2\tabcolsep
            \kern-\@tempdimb
            \leaders\vrule
                    \hskip\@tempdima\@plus  1fill
            \kern-\@tempdimc
            \hskip-\wd\z@ \@plus -1fill }
\makeatother
\newcommand{\expect}[1]{\langle #1 \rangle}

\newcommand{\smu}{Department of Physics,
Southern Methodist University, 
Dallas, TX 75275, USA}

\newcommand{\Dunlap}{Dunlap Institute for Astronomy \& Astrophysics,
University of Toronto,
Toronto, ON M5S 3H4, Canada}

\newcommand{\UofT}{David A. Dunlap Department of Astronomy \& Astrophysics, 
University of Toronto,
Toronto, ON M5S 3H4, Canada}

\newcommand{\ASU}{School of Earth and Space Exploration,
Arizona State University, 
Tempe, AZ 85287, USA}

\def\vl{\boldsymbol{\ell}}
\def\vlp{\boldsymbol{\ell}'}

\def\vL{\bm{\check{L}}}
\def\vLp{\bm{\check{L}}'}
\def\vd{\mathbf{d}}
\def\nhat{\hat{\mathbf{n}}}
\mathchardef\mhyphen="2D
\def\planck{\textit{Planck}}
\def\dd{{\rm d}}
\def\grad{\boldsymbol{\nabla}}
\def\Lcheck{{\check{L}}}
\def\Mcheck{{\check{M}}}

\begin{document}

\title{Small Correlated Against Large Estimator (SCALE) for Cosmic Microwave Background Lensing}

\author{Victor C. Chan}
\affiliation{\UofT}

\author{Ren\'ee Hlo\v{z}ek} 
\affiliation{\Dunlap}
\affiliation{\UofT}

\author{Joel Meyers}
\affiliation{\smu}

\author{Alexander van Engelen}
\affiliation{\ASU}

\begin{abstract}
    Weak gravitational lensing of the cosmic microwave background (CMB) carries imprints of the physics operating at redshifts much lower than that of recombination and serves as an important probe of cosmological structure formation, dark matter physics, and the mass of neutrinos.
    Reconstruction of the CMB lensing deflection field through use of quadratic estimators has proven successful with existing data but is known to be sub-optimal on small angular scales ($\ell > 3000$) for experiments with low noise levels.
    Future experiments will provide better observations in this regime, but these techniques will remain statistically limited by their approximations. We show that correlations between fluctuations of the large-scale temperature gradient power of the CMB sourced by $\ell < 2000$, and fluctuations to the local small-scale temperature power reveal a lensing signal which is prominent in even the real-space pixel statistics across a CMB temperature map. We present the development of the Small Correlated Against Large Estimator (SCALE), a novel estimator for the CMB lensing spectrum which offers promising complementary analysis alongside other reconstruction techniques in this regime. The SCALE method computes correlations between both the large/small-scale temperature gradient power in harmonic space, and it is able to quantitatively recover unbiased statistics of the CMB lensing field without the need for map-level reconstruction. SCALE can outperform quadratic estimator signal-to-noise by a factor of up to 1.5 in current and upcoming experiments for CMB lensing power spectra $C_{6000<L<8000}^{\phi\phi}$.
\end{abstract}

\maketitle

\section{Introduction}
Gravitational lensing of the cosmic microwave background (CMB) by cosmological structures along the line of sight has become a standard observational tool to probe the content and evolution of the universe; see Ref.~\cite{Lewis:2006fu} for a review. Improvements in measurements of the CMB with telescopes like the Atacama Cosmology Telescope (ACT,\cite{ACT:2020gnv}), the South Pole Telescope (SPT,\cite{SPT:2017jdf}) and the \planck~satellite \cite{Planck:2018nkj} have unveiled the fluctuations in the temperature and polarization signal of this primordial light down to arcminute scales. Gravitational lensing distorts our view of this primordial radiation, imparting non-stationary statistics that can be teased apart from the primordial fluctuations~\cite{SPT:2014puc,Planck:2018lbu,Darwish:2020fwf}.

Extracting (or `reconstructing') this signal from CMB temperature $T$ and/or polarization $E$ and $B$ maps can proceed via a number of approaches. The pioneering work of Refs.~\cite{Hu:2001kj,Okamoto:2003zw} developed the concept of the quadratic estimator (QE) for the lensing signal, which combines pairs of observed maps $TT$, $EE$, $BB$, $TE$, $TB$, and $EB$, to `reconstruct' the lensing potential field. Lensing reconstruction via the QE has been successful with existing data. The iterative $EB$ estimator~\cite{Hirata:2003ka,Smith:2010gu} is especially effective at large angular scales $(L \lesssim 1000)$ due to the transfer of power from $E$ modes into $B$ modes, the latter of which contains only a meager signal in the primary CMB caused by a possible epoch of cosmic inflation. Application of the QE on \planck\ data has allowed for a 40$\sigma$ detection of gravitational lensing \cite{Planck:2018lbu}. However, the effectiveness of the QE may soon be limited as experiments push to smaller scales and lower noise. Indeed, the recent analysis of a set of very deep SPTpol data showed improved results compared with the more standard QE approach~\cite{Millea:2020iuw}. The QE formalism, which approximates the full maximum likelihood estimate of the signal, is statistically sub-optimal on small angular scales and in low-noise regimes \cite{Hirata:2002jy,Hirata:2003ka,Smith:2010gu,Horowitz:2017iql,Carron:2017mqf,Hadzhiyska:2019cle}.

\begin{figure}
	\centering
	\includegraphics[width=0.9\columnwidth]{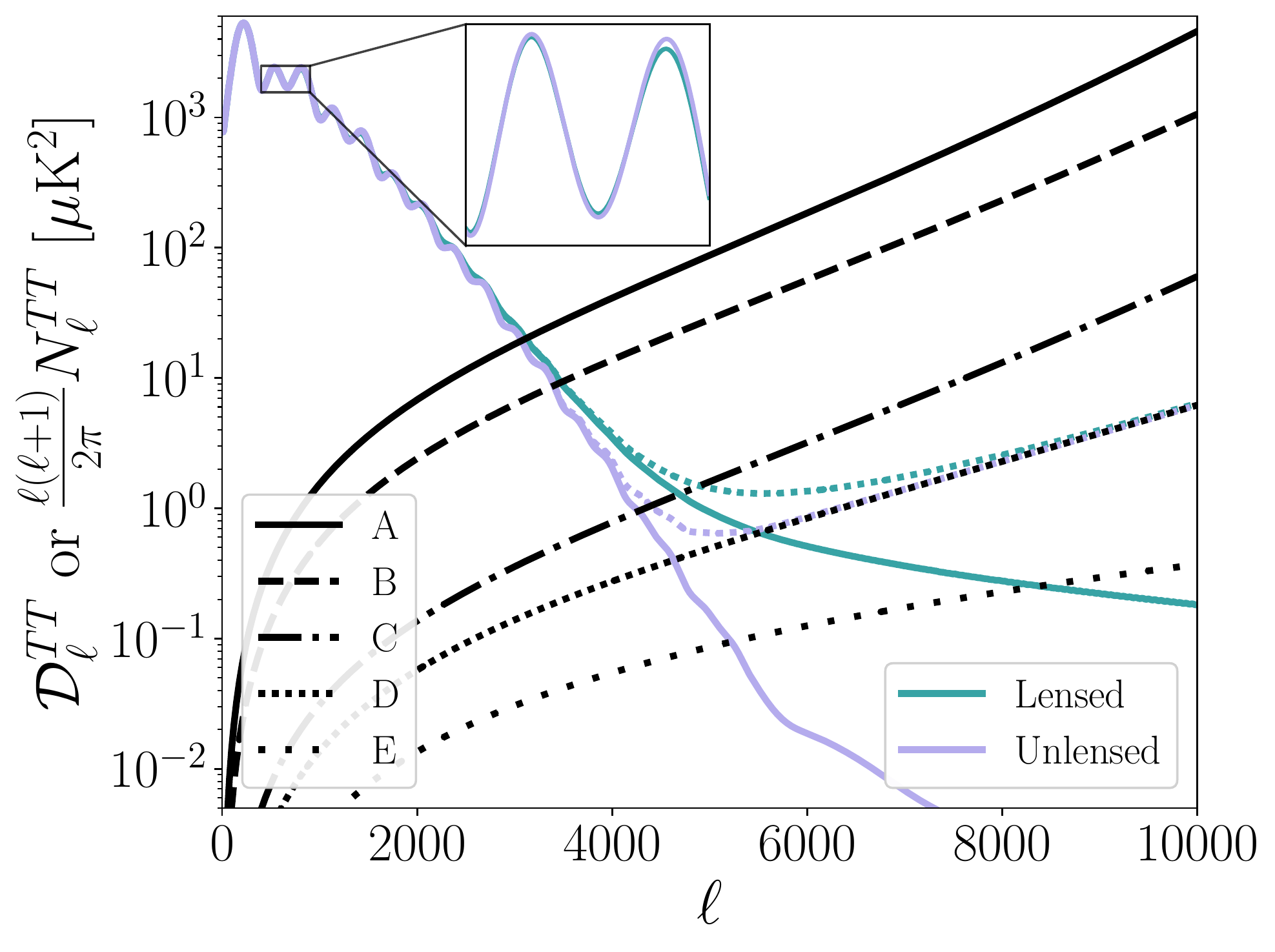}
	\includegraphics[width=0.9\columnwidth]{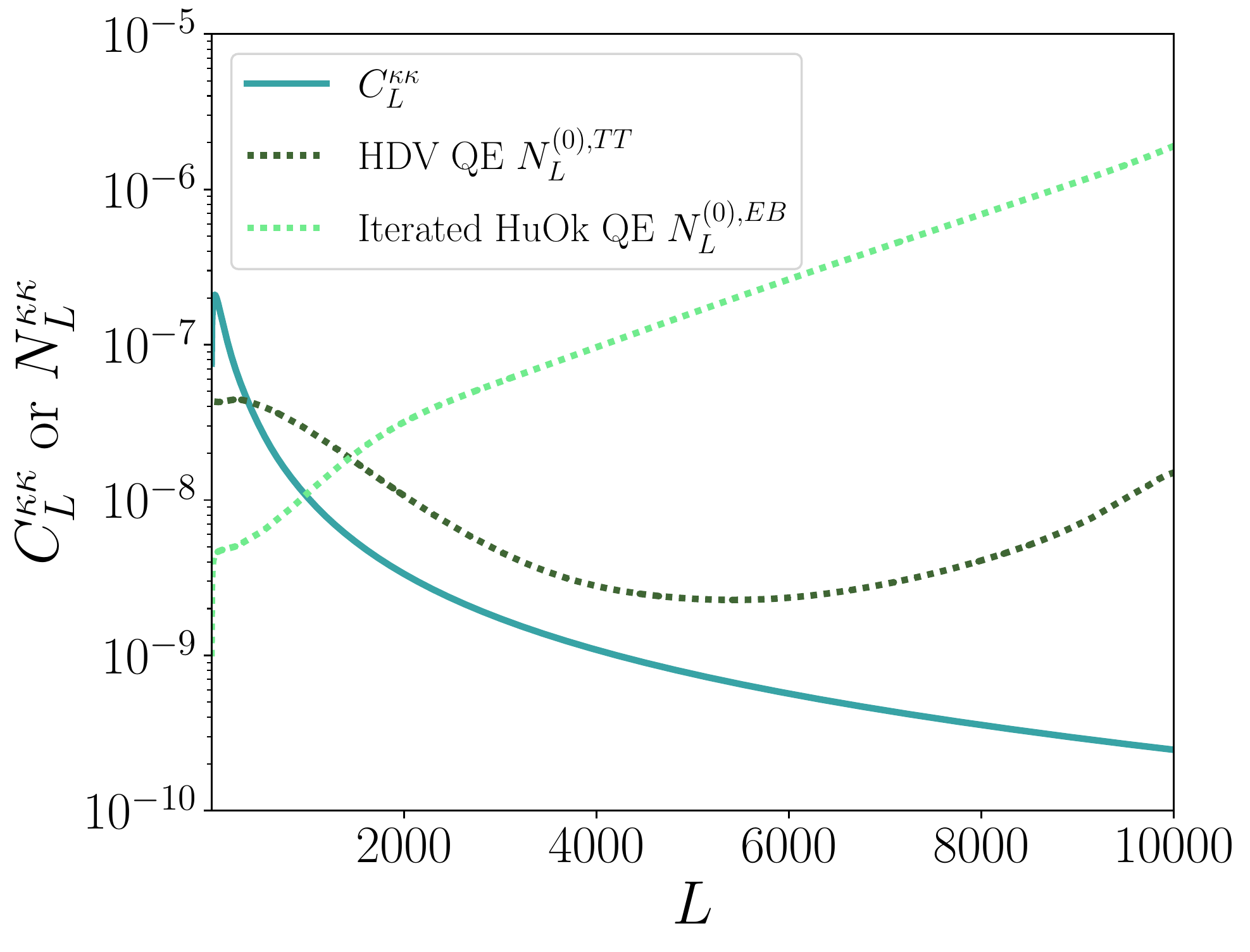}
	\caption{\textit{Top}: The input power spectrum for simulated raw CMB maps is shown in solid purple. The resulting lensed CMB power spectrum after applying a lensing field corresponding to the spectrum below is shown in solid teal. Noise spectra corresponding to different experiments are shown in black, and the lensed/unlensed spectra combined with configuration-D (defined below, in Table~\ref{tab:Noise}) noise are shown in their respective colours and dense dotted lines. \textit{Bottom}: The input lensing convergence power spectrum for simulated lensing potential fields. Also shown is the optimal reconstruction noise $N_L^{(0)}$ for the Hu, DeDeo \& Vale $TT$~\citep[HDV,][]{Hu:2007bt} and iterative Hu \& Okamoto $EB$ quadratic estimators~\citep[HuOk][]{Hu:2001kj,Okamoto:2003zw} computed with noise from configuration-D.
	}
	\label{fig:InputPS}
\end{figure}

The derivation of the QE procedure relies on the assumption that lensing is a weak effect, in the sense that it has only a small effect on the statistics of the CMB sky.
This is an appropriate approximation for most of the regimes in which the QE has historically been applied, but at lower noise levels, this approximation quickly begins to break down.  On small angular scales, at $\ell \gg 2000$, the power resulting from gravitational lensing dominates over the primordial unlensed power spectrum, as can be seen in Figure~\ref{fig:InputPS}. In this lensing-dominated regime, QE techniques are sub-optimal in the limit of low noise levels. Figure~\ref{fig:InputPS} also shows that the additive bias $N_L^{(0)}$ from the $TT$ reconstruction noise becomes lower than that of the iterative $EB$ estimator at small angular scales $(L \gtrsim 2000)$ due to increasing noise from the $EB$ estimator. The noise power is greater than the signal power at these scales; however, this bias can be removed with well-established methods \cite{Dvorkin:2009,Hanson:2011}. Bias from higher-order reconstruction noise $N_L^{(1)}$ also need to be estimated and removed at small angular scales \cite{Kesden:2002ku}.
Recent advances in computation and statistical methodology allow for the computation of the full maximum likelihood lensing map \cite{Hirata:2002jy,Carron:2017mqf,Millea:2020cpw} thereby surpassing the performance of QE techniques in simulated data.  

Quadratic estimators can also be shown to be sub-optimal in the small-scale, low-noise limit because they are weighted by the sky-averaged variance of the large-scale modes, despite precise measurement of the large-scale modes. Due to cosmic variance on large scales, this weighting contributes to excess variance in the lensing reconstruction. 
This limitation can be circumvented with the the so-called `Gradient Inversion' approach to reconstruction \cite{Seljak:1999zn,Horowitz:2017iql,Hadzhiyska:2019cle}  which, unlike the QE technique, is not limited by cosmic variance exhibited by the large scale temperature fluctuations. 

Quadratic estimators and other estimators like the gradient inversion estimator aim to reconstruct a map of the underlying lensing potential \textit{explicitly}. 
A reconstructed map of the lensing field is valuable for delensing~\cite{Kesden:2002ku,Knox:2002pe,Seljak:2003pn,Green:2016cjr,Hotinli:2021umk} and for cross-correlation with other maps of large scale structure~\citep[e.g.][]{Sherwin:2012mr,HerschelATLAS:2014txv,Liu:2015xfa,Schmittfull:2017ffw,Robertson:2020xom,Darwish:2020fwf,Baxter:2022enq, DES:2022xxr,Lin:2020sbb,Piccirilli:2022myi}, but the lensing power spectrum carries valuable information even without an associated map-level reconstruction. Phenomena which impact matter clustering can be constrained using measurements of the matter power spectrum, without requiring a map of overdensities.  Examples include the effects of neutrino mass and related quantities~\cite{Kaplinghat:2003bh,Lesgourgues:2012uu,Green:2021xzn,Abazajian:2022ofy}, dark matter interactions~\cite{Tulin:2017ara,Gluscevic:2019yal,Buen-Abad:2021mvc}, ultralight dark matter~\cite{Hui:2016ltb,Ferreira:2020fam}, warm dark matter~\cite{Drewes:2016upu}, and baryonic feedback~\cite{Chisari:2019tus}.

Our goal is to devise a simple estimator that can leverage the low-noise and high-resolution maps expected from future CMB surveys to measure the small-scale lensing power spectrum. We present the Small Correlated Against Large Estimator (SCALE), a new method of obtaining the small-scale lensing power constructed from the cross-correlation between maps of the local large-scale and small-scale temperature power. This estimator is complementary to reconstruction techniques aimed at estimating a map of the lensing potential. It is similar in spirit to the maximum likelihood, maximum a posteriori, Gradient Inversion, and Bayesian  techniques~\cite{Hirata:2002jy, Carron:2017mqf, Hadzhiyska:2019cle, Millea:2020cpw} in that it aims to make optimal use of lensing information at small-scales of a CMB temperature map. In contrast to the QE method, SCALE is designed to work on small angular scales, which leverages the ongoing improvements to detectors and telescopes in the coming decade \cite{CMB-S4:2016ple,SimonsObservatory:2018koc,NASAPICO:2019thw,Abazajian:2019eic,Sehgal:2019ewc}. SCALE specifically aims to avoid the extra variance incurred by QE techniques due to cosmic variance of the large-scale CMB temperature gradient, while also circumventing the highly correlated nature of QE errors at small angular scales. In contrast to the gradient inversion method described above, the SCALE pipeline consists of high-pass and low-pass filtered maps which are squared and then cross-correlated to estimate the lensing power spectrum directly, rather than a map of the lensing potential. 

We start with a brief review of CMB lensing in Section~\ref{sec:method/intro} and develop a simple test in real space to illustrate the principles of our proposed method in Section~\ref{sec:method/real}. We further develop this method, and present the SCALE procedure in Section~\ref{sec:method/ell}. After introducing our data simulations in Section~\ref{sec:Sims}, we present our results in Section~\ref{sec:results} and conclude in \ref{sec:discuss}.

%%%%%%%%%%%%%%%%%%% C_\ell^TT and C_\ell^kk %%%%%%%%%%%%%%%
%%%%%%%%%%%%%%%%%%%%%%

\section{Review of CMB lensing} \label{sec:method/intro}

In the absence of foregrounds and noise, the observed CMB temperature field $\Tilde{T}$ at a particular line of sight $\nhat$ is the unlensed temperature $T$ at a lensing deflection angle $ \vd(\nhat) $ away from the line of sight. The lensing deflection angle $ \vd = \grad\phi $ is the gradient of the lensing potential $ \phi $ when working within the Born approximation. We denote a gradient across the sky (i.e. along the plane perpendicular to the line of sight) with $\grad$. The lensed temperature is
\begin{equation}\label{eq:TaylorExpansion}
    \Tilde{T}(\nhat) = T(\nhat + \grad \phi(\nhat)) = T(\nhat) + \grad \phi(\nhat) \cdot \grad T(\nhat) + \dots \, .
\end{equation}
The lensing potential is directly related to the lensing convergence $\kappa=-\nabla^2\phi/2$, with power spectra related by $C_L^{\kappa\kappa} = (L(L+1))^2 C_L^{\phi\phi} / 4$. In our conventions, the Fourier transform of the temperature gradient is
\begin{equation}\label{eq:GradFourier}
    \grad T(\nhat) = i\int\frac{\dd^2\vl}{2\pi}\vl T(\vl) e^{i\vl\cdot\nhat} \, .
\end{equation}
Note that the majority of the CMB temperature gradient comes from modes with $\ell \lesssim 2000$ \cite{Hu:2007bt}. Taking Eq.~\eqref{eq:TaylorExpansion} into Fourier space, we apply Eq.~\eqref{eq:GradFourier} in combination with the convolution theorem to get 
\begin{align}
    \tilde{T}(\vl) &= \int\dd\nhat \tilde{T}(\nhat)e^{-i\vl\cdot\nhat} \nonumber\\
    &= T(\vl) - \int\frac{\dd^2\vlp}{2\pi}\vlp\cdot(\vl-\vlp)\phi(\vl-\vlp)T(\vlp) + \mathcal{O}(\phi^2) \label{eq:TFourierFull}
\end{align}
In Eq.~\eqref{eq:TFourierFull}, we see that at first order in $\phi$ the lensed temperature field $\tilde{T}$ is a convolution between the lensing potential field $\phi$ and the original unlensed temperature field $T$. Taking the two-point auto-correlation of the temperature field (e.g., steps (4.7) - (4.11) in Ref.~\cite{Lewis:2006fu}) yields the lensed power spectrum:
\begin{align}
\label{eq:LensedCMBPowerFull}
    \tilde{C}_\ell^{TT} \approx& \bigg(1 - \int\frac{\dd^2\vlp}{(2\pi)^2}C_{\ell'}^{\phi\phi} (\vl\cdot\vlp)^2 \bigg)C_\ell^{TT}  \nonumber \\
    &+ \int\frac{\dd^2\vlp}{(2\pi)^2}[\vlp\cdot(\vl-\vlp)]^2 C_{\ell'}^{TT} C_{|\vl-\vlp|}^{\phi\phi} \, .
\end{align}
Note that the first term is a cross-term between the zeroth and second order terms of Eq.~\eqref{eq:TFourierFull}, and the second term is a product of the first order term with itself. Similar to Eq.~\eqref{eq:TFourierFull}, it expresses that the lensed CMB temperature power contains a convolution between the lensing potential power and the original CMB temperature power. 
The effects of weak gravitational lensing on the CMB do not add or remove from the total CMB temperature variance $\int\dd\ell\ell C_\ell^{TT} /2\pi$ across the sky. Instead, lensing serves to redistribute power $C_\ell^{TT}$ between angular modes $\ell$ in a way that ``smooths out" the peaks and troughs in the observed power spectrum (as can be seen in Figure~\ref{fig:InputPS}); the power redistributed to scales $\ell \gtrsim 4000$ dominates the signal compared to the unlensed temperature modes which are suppressed by diffusion damping. Traditional estimators of the lensing potential take advantage of the correlations between angular modes that have been introduced, and they work to reconstruct the lensing potential field through measurement of these off-diagonal couplings.
We can make approximations to simplify Eq.~\eqref{eq:LensedCMBPowerFull} in the small-scale limit $\ell \gg 2000$.
The CMB temperature gradient variance, which we denote $\expect{|\grad T_L|^2}$, is made up of an integral over the larger scale modes $C_{\ell \lesssim 2000}^{TT}$ of the original CMB temperature field, given by $\int\dd\ell\ell^2 C_\ell^{TT} / 2\pi$. This background temperature gradient is approximately constant at small scales, which can be enforced with $\ell'\ll\ell$ in Eq.~\eqref{eq:LensedCMBPowerFull}. One can apply these approximations to arrive at a simplified representation of the lensed CMB temperature power on small scales (e.g., \S4.1.3 of Ref.~\cite{Lewis:2006fu}):
\begin{align}
    \tilde{C}_{\ell\gg2000}^{TT} &\approx \ell^2 C_\ell^{\phi\phi}\int\frac{\dd\ell'}{\ell'}\frac{\ell'^4C_{\ell'}^{TT}}{4\pi} + C_{\ell,r}^{TT} \nonumber\\
    &= \frac{1}{2}\expect{|\grad T_L|^2}\ell^2 C_\ell^{\phi\phi} + C_{\ell,r}^{TT}.\label{eq:SSClTT}
\end{align}
Here, we define $ C_{\ell,r}^{TT} $ which represents all remaining contributions to the observed temperature power which are not expected to strongly correlate with the large-scale gradient of the CMB temperature field. This includes the first term of Eq.~\eqref{eq:LensedCMBPowerFull} which contains a small amount of the unlensed CMB temperature power suppressed by diffusion damping crossed with a second-order lensing contribution. We may also include contributions from instrument noise, foregrounds, and other secondaries in $ C_{\ell,r}^{TT} $. 

A straightforward method to estimate the small-scale lensing power is to simply divide the observed excess small-scale temperature power by the average unlensed temperature gradient power on large scales. That is, we can rework Eq.~\eqref{eq:SSClTT} and estimate the small scale lensing power spectrum as
\begin{equation}
    C_\ell^{\phi\phi} \approx \frac{\tilde{C}_\ell^{TT}-C_{\ell,r}^{TT}}{\ell^2\left(\frac{1}{2} \left\langle |\grad T|^2 \right\rangle \right) } \, ,
\label{eq:Small_Scale_Lensing_Estimate}    
\end{equation}
for $\ell \gg 2000$. The motivation for our SCALE technique is that we can do better than Eq.~\eqref{eq:Small_Scale_Lensing_Estimate}, even without reconstructing a map of the lensing field. In any given patch of sky, the large-scale temperature gradient power around the line-of-sight $\nhat$ will deviate from the sky average due to random fluctuations.  As a consequence, the \textit{local} small-scale temperature power that results from lensing will also deviate from the sky average.  By correlating the spatial variations in the locally measured large-scale temperature gradient power with the spatial variations in the small-scale temperature power, we can construct an improved estimate of the small-scale lensing power.  Furthermore, variations in the observed small-scale temperature power that are due to sources other than lensing (such as non-stationary noise or astrophysical foregrounds), are not expected to correlate with variations in the large-scale temperature gradient power since these effects result from survey choices or local physics unrelated to the long wavelength fluctuations responsible for the large-scale temperature gradients.

In summary, we propose a new lensing estimator with a similar form to Eq.~\eqref{eq:Small_Scale_Lensing_Estimate}. The key difference is allowing the \textit{local} small-scale lensed temperature power to fluctuate according to the steepness of the background temperature gradient in the same part of the sky:
\begin{align}
    \tilde{C}_{\ell\gg2000}^{TT,\mathrm{local}}(\nhat) &\approx \frac{1}{2}{|\grad T_L(\nhat)|^2}\,\ell^2 C_\ell^{\phi\phi} + C_{\ell,r}^{TT}.\label{eq:localSSClTT}
\end{align}
Eq.~\eqref{eq:Small_Scale_Lensing_Estimate} is recovered by taking the sky average of this version. We consider combinations of CMB temperature maps because temperature-based lensing reconstruction out-performs polarization-based estimators on small angular scales, due to the fact that polarization maps become dominated by noise at these angular scales (see Figure~\ref{fig:InputPS}). The lensing estimator we propose shares some similarities with techniques used to measure the kinetic Sunyaev-Zel'dovich (kSZ) effect through variations in small-scale temperature power~\cite{Smith:2016lnt}.

In the following section, we present a simple proof-of-concept that takes advantage of the \textbf{local map-space correlations between the small-scale temperature power and the square of the observed temperature gradient amplitude} in order to tease out the statistics of the underlying lensing potential field.  

\section{Introductory concepts in Real Space \label{sec:method/real}}
Before describing the harmonic-space implementation of SCALE in this work, we start with an illustrative real-space description of the concept to provide the intuition and motivation for the techniques we develop in the following section. We introduce the notation $\overline{X}(\nhat)$ along with shorthand $\overline{X}$ to indicate the \textit{local} average of a quantity $X$ near a line-of-sight $\nhat$.

The total observed small-scale CMB power along a given line of sight $ \overline{T_S^2}(\nhat) $ is given by the 2-dimensional (2D) angular integral over Eq.~\eqref{eq:localSSClTT}. We also specify the large-scale temperature gradient amplitude $\overline{|\grad T_L(\nhat)|}$ to still be approximately constant near each single line of sight while allowing for the small fluctuations between different lines of sight as described in the previous section. Since we do not expect the lensing potential field to strongly correlate with the remaining contributions to the small-scale temperature power, we generally expect
\begin{equation}\label{eq:Line}
    \overline{T_S^2}(\nhat) \simeq a_1\overline{|\grad T_L(\nhat)|}^2 + a_0,
\end{equation}
where we have the following contributions:
\begin{enumerate}
	\item A term containing the lensing contribution which scales with the amplitude of the large-scale CMB temperature gradient $\overline{|\grad T_L(\nhat)|}^2 $, and
	\item a term containing all remaining contributions which \textit{do not} scale with the CMB temperature gradient.
\end{enumerate}
 Eq.~\eqref{eq:Line} motivates a simple, map-space approach to gather local statistics for $\overline{T_S^2}(\nhat)$ and $\overline{|\grad T_L(\nhat)|}^2$ and to take advantage of their correlations to bring out the lensing signal. The expectation is that $a_1 \rightarrow 0$ in a CMB temperature map \textit{without} lensing, and $a_1$ should increase with a stronger lensing signal (i.e., $a_1 \propto C_L^{\phi\phi}$, c.f., \ref{eq:localSSClTT}.). Contributions to the temperature map such as noise and foregrounds that do not come from lensing  should directly contribute to $a_0$, but not $a_1$ because they are not expected to correlate with the large-scale CMB temperature gradient.  We should therefore be able to infer the small-scale lensing power from the measured value of $a_1$.

\subsection{Map reduction to local patches}

Of the two observable quantities in Eq.~\eqref{eq:Line}, we begin with measuring background temperature gradient $ \overline{\grad T_L} $ as well as the small-scale temperature $T_S$ from a single input CMB temperature map. Local statistics of $ \overline{|\grad T_L|}^2 $ and $\overline{T_S^2}$ then need to be gathered in small cutouts of the observed field. 

To ensure that we are only including the smooth component of the temperature gradient, we filter the maps in Fourier space. We compute maps of the observed CMB temperature gradient using
\begin{equation}\label{eq:Tgradient}
    \grad T_L(\vl) = i\vl T(\vl).
\end{equation}
We apply a low-pass top hat filter before returning the map to pixel space: $ \grad T_L(\ell>\ell_{\nabla T}) = 0 $.

It is not immediately obvious what scale $ \ell_{\nabla T} $ should be used in the low-pass filter. Enough modes should be included such that the resulting $ \grad T_L $ maps contain enough information about the background temperature gradient, and $ \ell_{\nabla T} $ should be small enough such that there are no direct correlations from similar shared modes between $ \grad T_L $ and the small scale temperature $ T (\ell \gg 2000) $. The large-scale CMB temperature gradient $ \grad T_L| $ that we are looking for is mostly constituted by modes $ \ell \lesssim 2000 $, so we consider a low-pass cutoff at $ \ell_{\nabla T} = 3000 $. It is important to keep in mind that pixels near the boundary of the $ \grad T_L $ map may be unusable if the original $ T $ map does not have the appropriate repeating boundary conditions. 

%%%%%%%%%%%%%% Patches Flowchart %%%%%%%%%%%%%%%%
\begin{figure}
    \centering
    \includegraphics[width=0.9\columnwidth]{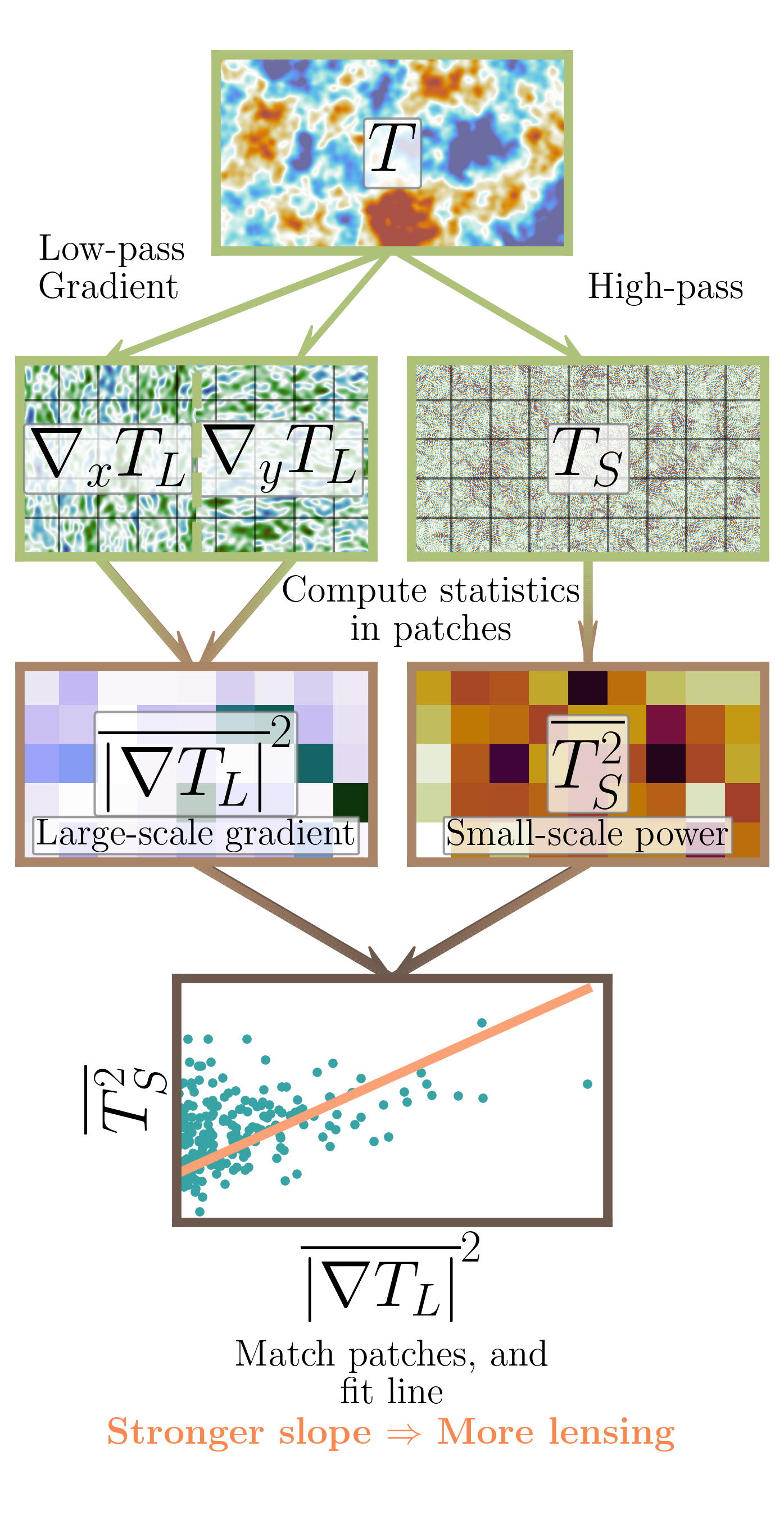}
    \caption{Schematic of a procedure to reduce a CMB temperature map to local, real-space statistics which can then be correlated to infer lensing effects. Maps in the original resolution are denoted in \textit{green} borders, and the \textit{light brown} borders indicate a degraded resolution after computing relevant statistics within local patches of width $40'$. Note that the patches shown here are for visual presentation, and they are larger than those chosen later on.}
    \label{fig:PatchesFlowchart}
\end{figure}
%%%%%%%%%%%%%%%%%%%%%%%%%%%%

In order to pick out only the small scale lensing power contributions which we wish to correlate with the background temperature gradient, the original CMB temperature map also needs to be filtered, this time with a high-pass window ($ \ell_{\rm min} < \ell < \ell_{\rm max} $) for the scales relevant to the analysis. A variety of choices for $ \ell_{\rm min} $ and $ \ell_{\rm max} $ can be made as long as scales are in a regime where lensing is dominant ($ \ell \gg 2000 $), and we require $\ell_{\rm min} > \ell_{\nabla T}$ to ensure that any correlations between the large-scale gradient and small-scale power fields is strictly from lensing. The choice in $ \ell_{\rm min} $ and $ \ell_{\rm max} $ affect the expected $ a_1 $ and $ a_0 $ in Eq.~\eqref{eq:Line} through the angular integral for the total CMB temperature power.

We then split both filtered maps into patches in order to estimate the quantities $ \overline{|\grad T_L|}^2 $ and $ \overline{T_S^2} $ along different lines of sight $ \nhat $. The required patch size is also not obvious, but needs to be small such that the assumption of $ \overline{|\grad T_L|}^2 $ being constant within a patch is reasonable. In considering the characteristic angular scale $ \theta \sim 2\pi / \ell $ for $ \ell = 2000 $, the patches should be $ \lesssim 10' $ wide. The patches also need to be large enough such that there are sufficient pixels of the map within each patch to make good estimates of $\overline{|\grad T_L|}^2$ and $\overline{T_S^2}$. This is also dependent on the resolution of the original map, which will be discussed further in \S~\ref{sec:Sims}. For the example shown here, we choose a patch size of $10' \times 10'$, which is $20 \times 20$ pixels in our simulated maps with resolution $0.5'$.

%%%%%%%%%%%%%% Scatter Plot Line Fit %%%%%%%%%%%%%%%%%%%%%%%%%%%%%%
\begin{figure*}
    \centering
    \includegraphics[width=1.8\columnwidth]{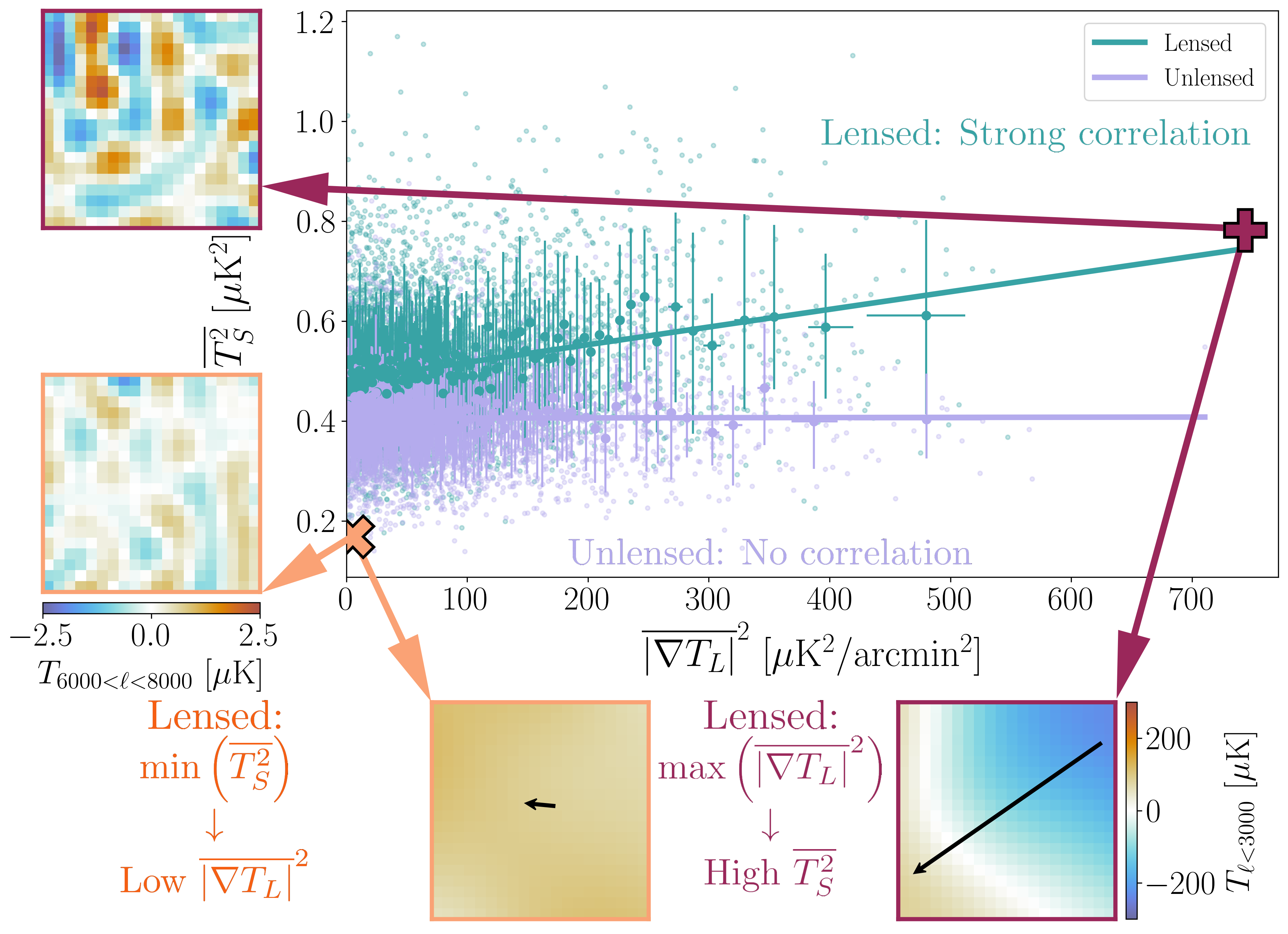}
    \caption{\textit{Main}: Local small scale ($6000 < \ell < 8000$) temperature variance $ \overline{T_S^2} $ vs. average large scale ($\ell < 3000$) temperature gradient amplitude squared $ \overline{|\grad T_L|}^2 $ for lensed (\textit{teal}) and unlensed (\textit{purple}) realizations are shown here as faint, small points. A low-variance/small-gradient patch from the lensed realization (\textit{orange} x), and a high-variance/large-gradient patch from the lensed realization (\textit{maroon} +) are highlighted. The larger points are centered on the medians within bins of $ \overline{|\grad T_L|}^2 $ containing an equal number of patches, with error bars corresponding to $68\%$ quantiles. The lines of best fit through the binned points are also shown. \textit{Left}: $10'\times10'$ cutouts of the lensed CMB temperature map filtered for small scales ($6000 < \ell < 8000$) corresponding to the highlighted patches. \textit{Bottom}: $10'\times10'$ cutouts of the lensed CMB temperature map filtered for large scales ($\ell < 3000$) corresponding to the highlighted patches. The average gradient direction and relative amplitudes across each patch are shown with the overlaid arrows.}
    \label{fig:ScatterPlotPatches}
\end{figure*}
%%%%%%%%%%%%%%%%%%%%%%%%%%%%%%%%%%%

The quantity $ \overline{|\grad T_L(\nhat)|}^2 $ can now be computed for every patch across the map. For each of the two perpendicular directions on the map $ \hat{x} $ and $ \hat{y} $, we first compute the average gradient across each patch $ \overline{\grad T_L}(\nhat) =  \overline{\nabla_x T_L}\hat{x} + \overline{\nabla_y T_L}\hat{y} $. We can then readily compute $ \overline{|\grad T_L(\nhat)|}^2 = (\overline{\nabla_x T_L})^2 + (\overline{\nabla_y T_L})^2  $ for each patch.

\subsection{Lensing from patch statistics}

The remaining quantity to compute within each patch is $ \overline{T_S^2}(\nhat) $. We compute the auto-variance of the high-pass filtered $T_S$ map within each patch. The $\overline{|\grad T_L|}^2$ and $\overline{T_S^2}$ from each patch, in principle, provides a very noisy and approximate estimate of the overall small-scale lensing power present in the map, following a distribution about Eq.~\eqref{eq:Line}. By combining the statistics of many noisy patches across the map, there is opportunity to more rigorously quantify the slope in Eq.~\eqref{eq:Line}, and relate it to the 2D-angular integral of Eq.~\eqref{eq:SSClTT} in the appropriate space of chosen $\ell_{\nabla T}$, $\ell_{\rm min}$, and $\ell_{\rm max}$. 

The outputs of this real-space procedure applied to both a lensed and unlensed realization of the CMB temperature are shown in Figure~\ref{fig:ScatterPlotPatches} to illustrate the relationship between them. A positive correlation between $\overline{T_S^2}$ and $\overline{|\grad T_L|}^2$ can be clearly seen in the ensemble of lensed patches, while no significant correlation is seen in the sample of unlensed patches. Two example patches are also highlighted in Figure~\ref{fig:ScatterPlotPatches}, and the corresponding cutouts of the lensed CMB temperature filtered to the relevant scales of the small-scale temperature and the large-scale gradient are shown. The cutouts show that, even upon visual inspection, typical areas on the lensed CMB with a steep background temperature gradient usually have a higher small scale temperature power than typical areas with a relatively weak background temperature gradient.

There are several challenges that must be overcome in order to use this real-space method as a reliable estimator of CMB lensing. First, it is important to note that the strictly positive nature of the auto-variance forces the observed distribution of $ \overline{T_S^2}(\nhat) $ across the map to be positively skewed. It is also non-trivial to determine a choice in patch size that optimally includes as many pixels per patch while keeping the large-scale gradient and lensing statistics consistent within each patch. In fact, we show in Section~\ref{sec:method/ell} and Figure~\ref{fig:Lcheck} that there is no single patch size which can be chosen to effectively capture all the correlations between the large and small scale temperature fluctuations. These factors, combined with the fact that the temperatures observed in neighboring real-space pixels across each patch are highly correlated, suggests that the expected distribution of observed $ \overline{T_S^2}(\nhat) $ about Eq.~\eqref{eq:Line} is non-trivial. One option to make this distribution better behaved is to compute the covariance of two observations of the same CMB temperature field. By splitting up time-ordered CMB observations into two or more maps of the same area of sky, one can take advantage of the fact that the maps contain the same CMB realization (which should contain the same lensing information and correlations) and different noise realizations (which should not co-vary across maps).

One more challenge with quantifying this method is the loss of information coming from the local large-scale gradient direction when computing $\overline{|\grad T_L(\nhat)|}^2$. One may choose to construct individual filters for each real-space patch and its observed gradient direction in order to focus on the expected lensing signal(s). One such example is choosing a filter $f_\ell = \cos\alpha = \hat{\nabla} \cdot \hat{\ell}$ in addition to the high-pass filter for the small-scale temperature. In practice, this means that the small-scale temperature patches must each be filtered separately and uniquely based on the observed gradient direction in each patch $\hat{\nabla} T_L (\nhat)$. This approach once again faces the previous challenge of the large-scale temperature gradient fluctuations not being fully represented within a single patch size. The non-Hermitian nature of such a filter also introduces edge-effects along the borders of each patch, which alters information from an already limited set of pixels within each patch.

While we limit our current presentation of this real-space procedure to a qualitative analysis, it provides significant intuition and motivation for the development of SCALE. Figure~\ref{fig:ScatterPlotPatches} demonstrates that the small-scale CMB temperature fluctuations are intricately tied to the statistics of the underlying lensing field as well as the large-scale temperature fluctuations of the CMB itself. In other words, information about the lensing field naturally comes out when correlating small-scale CMB temperature fluctuations to large-scale CMB temperature fluctuations. While the observed CMB temperature field is expected to be contaminated by foregrounds and noise, we do not expect such contributions to be strongly correlated between small and large scales. 
These properties of a lack of noise correlation and the direct lensing correlation between the large-scale gradient and small-scale temperature are central to SCALE. This estimator overcomes the weaknesses of the real-space method, and we will show that it provides a quantitative estimate of the underlying small-scale lensing statistics in the following section.

%%%%%%%%%%%%%% LambdaSigma %%%%%%%%%%%%%%%%%%%%%%%%%%%%%%%%
\begin{figure*}
    \centering
    \includegraphics[width=1.2\columnwidth]{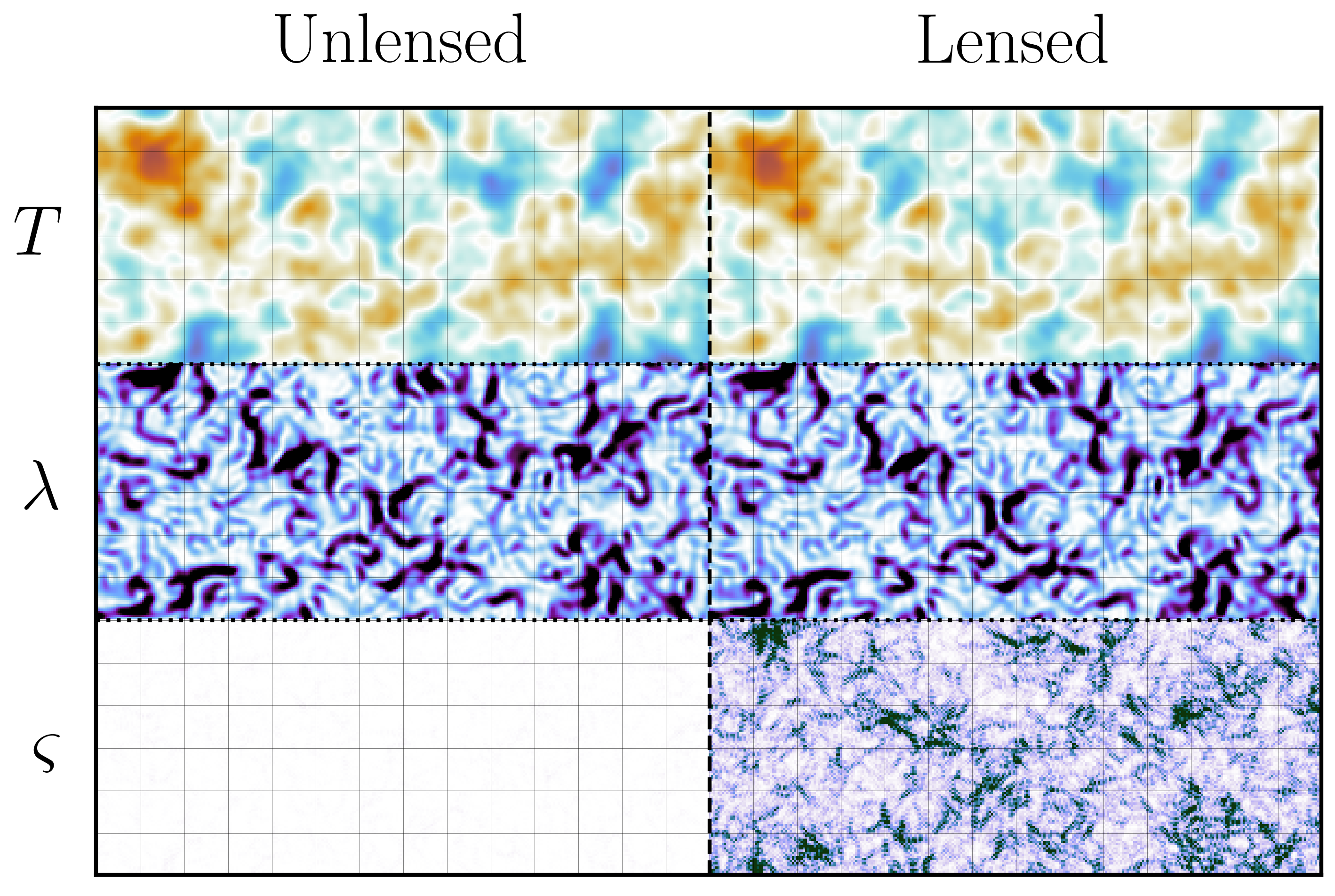}
    \caption{A comparison of a CMB temperature realization before and after lensing in the absence of noise and foregrounds. The same area of sky is shown for all panels, including a visualization of the $\lambda$ and $\varsigma$ maps derived from each version. Panels on the same row are shown with the same colormap and limits. A grid with $20'$ spacing is overlaid, which illustrates patches twice the width of the chosen patches for the real-space proof of concept.}
    \label{fig:LambdaSigma}
\end{figure*}
%%%%%%%%%%%%%%%%%%%%%%%%%%%%%%%%%%%%%%%%%%%%%%%%%%%%%%

%%%%%%%%%%%%%% Lcheck %%%%%%%%%%%%%%%%%%%%%%%%%%%%%%%%
\begin{figure*}
    \centering
    \includegraphics[width=1.6\columnwidth]{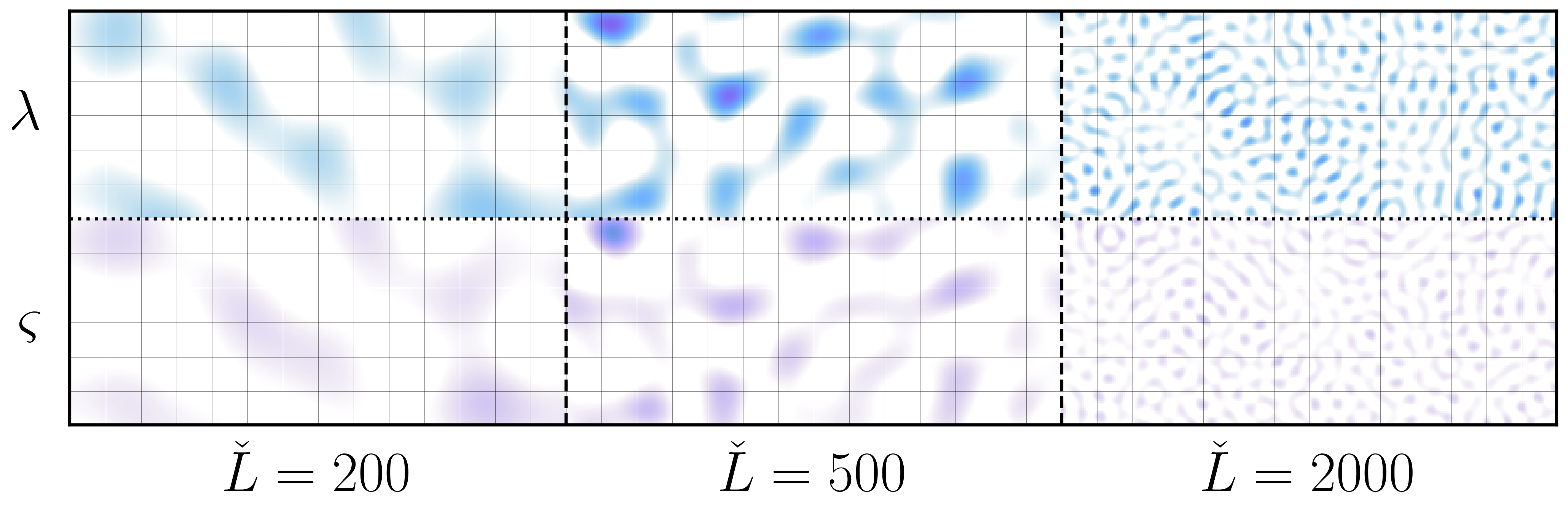}
    \caption{A visualization of several $\Lcheck$ bands of width $\Delta\Lcheck=300$ centered at the $\Lcheck$ labelled highlights the correlations between $\lambda$ and $\varsigma$ induced by lensing. The same lensed CMB realization from Figure~\ref{fig:LambdaSigma} is shown, with the same colormaps and limits for $\lambda$ and $\varsigma$. SCALE quantifies the correlations between the top and bottom panels in its estimates of the underlying lensing statistics. A grid with $20'$ spacing is overlaid, which illustrates patches twice the width of the chosen patches for the real-space proof of concept.}
    \label{fig:Lcheck}
\end{figure*}
%%%%%%%%%%%%%%%%%%%%%%%%%%%%%%%%%%%%%%%%%%%%%%%%%%%%%%

\section{The SCALE Method \label{sec:method/ell}}
The small-scale lensing signal is reflected in CMB temperature maps through local correlations between the small-scale temperature power fluctuations and the large-scale temperature power fluctuations across the sky. For a given map of the observed CMB temperature field $ T(\vl) $, we begin by constructing the fields containing the relevant information at small, and large scales. Here, we will sketch the procedure of forming the optimal direct estimate of the lens-induced correlation between small-scale gradient power and large-scale gradient power, in a way that minimizes the variance of the result.  The full details of the derivation are presented in Appendix \ref{sec:scalederivation}.

We begin by constructing \textit{large-scale} temperature gradient fields for two perpendicular directions on the map, $ \grad T_{L}(\vl) $, by applying the top hat filter $ W_\lambda(\vl) $ combined with a Wiener filter to the original temperature field,
\begin{align}
    W_\lambda(\vl) = 
    \begin{cases}
        1, & \ell_{2,\mathrm{min}} \leq |\vl| < \ell_{2,\mathrm{max}} \\
        0, & \text{else} \, ,
    \end{cases}
\end{align}
\begin{align}
    \grad T_{L}(\vl) = \frac{i\,\vl\,W_\lambda(\vl)\,C_\ell^{TT}\, T(\vl)}{C_\ell^{TT,\mathrm{obs}}}\, .\label{eq:GradTL}
\end{align}
Note that a fiducial temperature power spectrum $ C_\ell^{TT} $ is required for the Wiener filter in this step. It is not imperative that the assumed model exactly matches the underlying cosmology, as the results are not sensitive to this choice. The observed CMB temperature power spectrum, $ C_\ell^{TT,\mathrm{obs}} $, of the map is also required for our filters. We construct a field containing the large scale temperature power fluctuations after returning each gradient component to real space, squaring each component, and then adding them together:
\begin{align}
    \lambda(\nhat) = \big(\nabla_x T_L(\nhat)\big)^2 + \big(\nabla_y T_L(\nhat)\big)^2 \, .\label{eq:LambdaMap}
\end{align}
%

%%%%%%%%%%%%% SCALE Schematic %%%%%%%%%%%%%%%
\begin{figure}
    \centering
    \includegraphics[width=0.9\columnwidth]{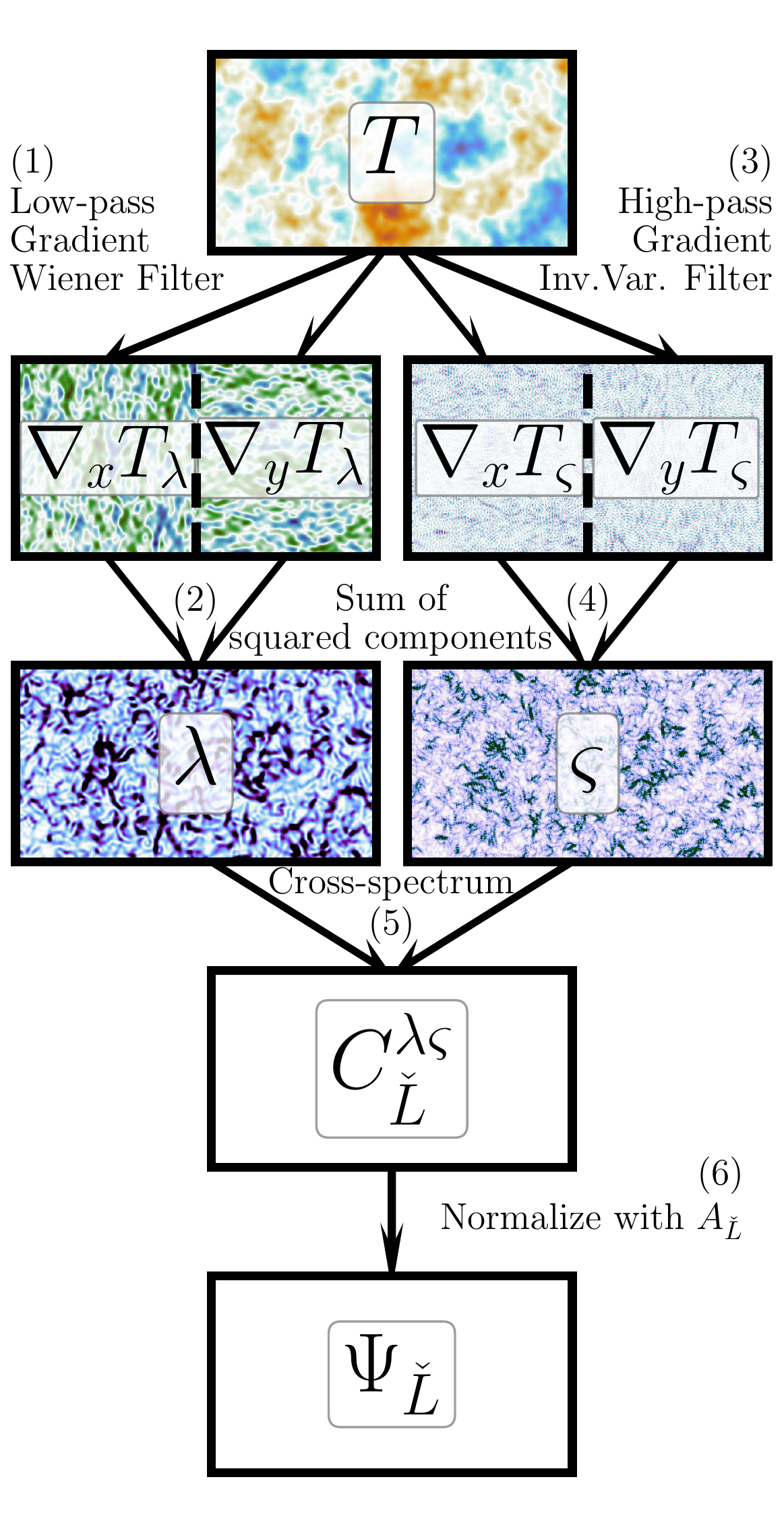}
    \caption{Schematic of the steps taken in SCALE pipeline.}
    \label{fig:SCALESchematic}
\end{figure}
%%%%%%%%%%%%%%%%%%%%%%

Similarly, we construct \textit{small-scale} temperature gradient fields in two perpendicular directions on the map, $ \grad T_{S}(\vl) $,  by applying a top hat filter $ W_\varsigma(\vl) $ combined with an inverse-variance filter to the observed temperature field,
\begin{align}
    W_\varsigma(\vl) = 
    \begin{cases}
        1, & \ell_{1,\mathrm{min}} \leq |\vl| < \ell_{1,\mathrm{max}} \\
        0, & \text{else} \, ,
    \end{cases}
\end{align}
\begin{align}
    \grad T_{S}(\vl) = \frac{i\,\vl\,W_\varsigma(\vl) \,T(\vl)}{C_\ell^{TT,\mathrm{obs}}} \, . \label{eq:GradTS}
\end{align}
We construct a field containing the small-scale temperature power fluctuations after returning each gradient component to real space, squaring each component, and then adding them together:
\begin{align}
    \varsigma(\nhat) = \big(\nabla_x T_S(\nhat)\big)^2 + \big(\nabla_y T_S(\nhat)\big)^2\, .\label{eq:SigmaMap}
\end{align}

The field $ \varsigma $ strictly contains the small-scale temperature power fluctuations at the scales allowed by the filter $ W_\varsigma(\vl) $. We expect this field to correlate with the large-scale temperature power fluctuations captured by the field $ \lambda $. Such correlations are only expected as a result of lensing on the original CMB temperature field because the filters are chosen to have disjoint support in $\ell$. As a result the cross-spectrum between each field, $ C_\Lcheck^{\varsigma\lambda} $, is a four-point function that estimates the power of the lensing potential $ C_L^{\phi\phi} $. Each mode $\Lcheck$ of the cross-spectrum represents a particular scale over which the fields $\lambda$ and $\varsigma$ correlate. This is illustrated in Figures~\ref{fig:LambdaSigma} and \ref{fig:Lcheck}. The $\lambda$ fields look visually similar between unlensed and lensed realizations of the CMB, but in the absence of noise, the $\varsigma$ fields show much stronger fluctuations in the lensed realization (unlensed small scale power is suppressed by diffusion damping). Fluctuations in the lensed $\varsigma$ field visibly correlate with the $\lambda$ field. Further filtering the $\lambda$ and $\varsigma$ fields illustrates the $\Lcheck$ modes probed by the cross-spectrum, and the lensing-induced correlation between $\lambda$ and $\varsigma$ becomes striking. 

Figure~\ref{fig:Lcheck} also visualizes how the real-space method from Section \ref{sec:method/real} was combining the information from many modes of $\Lcheck \gtrsim 500$ within each patch, and was including information from modes $\Lcheck \lesssim 500$ when performing the fit with many patches.

The introduction of noise (and foregrounds) adds power to both $\lambda$ and $\varsigma$ fields, and it can become the dominant source of power in the $\varsigma$ field. Noise contributions to $\lambda$ and $\varsigma$ are not expected to correlate with each other, meaning the cross-spectrum $C_\Lcheck^{\varsigma\lambda}$ is expected to be largely insensitive to noise (though noise will contribute to its variance).

For the SCALE cross-spectrum to be an unbiased estimate of the lensing power, it needs to be normalized to take into account the filtering that was applied, as well as the expected action of lensing on the fields:
\begin{align}
    \Psi_\Lcheck = A_\Lcheck C_\Lcheck^{\varsigma\lambda} \, .
\end{align}
The normalization $A_\Lcheck$ is computed as a double integral of both the observed temperature power spectrum $ C_\ell^{TT,\mathrm{obs}} $, and the fiducial temperature power spectrum $ C_\ell^{TT} $ used in the Wiener filter above:
\begin{align}\label{eq:ALDef1}
    &A_\Lcheck = \Bigg[ 2 \int \frac{\dd^2\vl_1}{(2\pi)^2} W_\varsigma(\vl_1)W_\varsigma(\vL-\vl_1)
    \nonumber \\
    & \qquad\qquad \times \left(\vl_1 \cdot (\vl_1-\vL) \right) \frac{1}{C_{\ell_1}^{TT,\mathrm{obs}}} \frac{1}{C_{|\vL-\vl_1|}^{TT,\mathrm{obs}}}   
    \nonumber \\
    & \qquad\qquad \times 
    \int \frac{\dd^2\vl_2}{(2\pi)^2} W_\lambda(\vl_2)W_\lambda(\vL-\vl_2)
    \left(\vl_2 \cdot (\vl_2-\vl_1) \right) \nonumber \\
    & \qquad\qquad\quad \times \left( (\vL-\vl_2) \cdot (\vl_1-\vl_2) \right) \left(\vl_2 \cdot (\vl_2-\vL) \right) \nonumber \\
    & \qquad\qquad\quad \times \frac{\left( C_{\ell_2}^{TT} \right)^2}{C_{\ell_2}^{TT,\mathrm{obs}}} \frac{\left( C_{|\vL-\vl_2|}^{TT} \right)^2} {C_{|\vL-\vl_2|}^{TT,\mathrm{obs}}} \Bigg]^{-1} \, , 
\end{align}
The bounds of each integral correspond to the scales allowed by the small-scale window function $W_\varsigma$, and the large-scale window function $W_\lambda$. See Appendix~\ref{sec:SCALEDerivation} for the steps leading to the definition of $A_\Lcheck$ in Eq.~\eqref{eq:ALDef}. The expected value of $ \expect{\Psi_\Lcheck} $ can be similarly computed with the lensing power $ C_\ell^{\phi\phi} $:
\begin{align}\label{eq:PsiExpected1}
    &\left\langle \Psi_\Lcheck \right\rangle = 2 A_\Lcheck
    \int \frac{\dd^2\vl_1}{(2\pi)^2} W_\varsigma(\vl_1)W_\varsigma(\vL-\vl_1) \nonumber \\
    & \qquad\qquad \times \left(\vl_1 \cdot (\vl_1-\vL) \right) \frac{1}{C_{\ell_1}^{TT,\mathrm{obs}}} \frac{1}{C_{|\vL-\vl_1|}^{TT,\mathrm{obs}}} \nonumber \\
    & \qquad\qquad \times 
    \int \frac{\dd^2\vl_2}{(2\pi)^2} W_\lambda(\vl_2)W_\lambda(\vL-\vl_2)
    \left(\vl_2 \cdot (\vl_2-\vl_1) \right)
     \nonumber \\
    & \qquad\qquad\quad \times \left( (\vL-\vl_2) \cdot (\vl_1-\vl_2) \right)
    \left(\vl_2 \cdot (\vl_2-\vL) \right) \nonumber \\
    & \qquad\qquad\quad \times \frac{\left( C_{\ell_2}^{TT} \right)^2}{C_{\ell_2}^{TT,\mathrm{obs}}} \frac{\left( C_{|\vL-\vl_2|}^{TT} \right)^2} {C_{|\vL-\vl_2|}^{TT,\mathrm{obs}}} C_{|\vl_1-\vl_2|}^{\phi\phi} \, .
\end{align}
Note that unlike for, e.g., the QE estimator, with the SCALE estimator we do not directly recover the signal of immediate interest (namely, $C_\ell^{\phi\phi}$ in this case).  This means that non-trivial physical changes to $C_\ell^{\phi\phi}$, such as from extensions to the cosmological model, would appear in the $\Psi_\Lcheck$ statistic in the SCALE estimator only indirectly, via this integral relation.  Nevertheless, we will show below that the expected $\Psi_\Lcheck$ is readily computed for any cosmological model, and shows excellent agreement with simulated reconstructions.

The expected noise variance of $\Psi_\Lcheck$, i.e., the variance in the absence of any lensing, is $N_\Lcheck \approx 4A_\Lcheck$, and the expected minimum uncertainty on an estimated $\hat{\Psi}_\Lcheck$ in the limit of low covariance between $\Lcheck$ modes is
\begin{align}\label{eq:PsiLError}
    \Delta\hat{\Psi}_\Lcheck = \sqrt{\frac{\Psi_\Lcheck^2+4A_\Lcheck}{f_\mathrm{sky}\Delta\Lcheck(2\Lcheck+1)}}.
\end{align}
The details of $A_\Lcheck$, $ \expect{\Psi_\Lcheck} $ and its expected variances are derived in Appendix~\ref{sec:SCALEDerivation} leading up to Eq.~\eqref{eq:PsiExpected}. We also demonstrate that the inverse-variance and Wiener filters are the optimal filters to minimize the noise variance $N_\Lcheck$ of the SCALE estimator.  

The general flow of the SCALE pipeline is illustrated in Figure~\ref{fig:SCALESchematic}, and summarized here beginning with a temperature map $T(\nhat)$: 
\begin{enumerate}
    \item Transform $T(\nhat)$ into harmonic space $T(\vl)$, apply the operations described by Eq.~\eqref{eq:GradTL}, and return $\grad T_L$ to map space.
    \item Compute $\lambda(\nhat)$ using Eq.~\eqref{eq:LambdaMap}.
    \item Apply the operations described by Eq.~\eqref{eq:GradTS} to $T(\vl)$, and return $\grad T_S$ to map space.
    \item Compute $\varsigma(\nhat)$ using Eq.~\eqref{eq:SigmaMap}.
    \item Compute the cross-spectrum $C_\Lcheck^{\lambda\varsigma}$.
    \item Apply the normalization $\Psi_\Lcheck = A_\Lcheck C_\Lcheck^{\lambda\varsigma}$.
\end{enumerate}
The end result $ \Psi_\Lcheck $ is a set of separate estimates of the lensing power spectrum $ C_\ell^{\phi\phi} $ weighted by the normalization $ A_\Lcheck $ along a range of scales set by $ W_\varsigma(\vl) $ and $ W_\lambda(\vl) $.
We note that the nature of our effectively four-point correlator is reminiscent of the trispectrum calculations for ${N}^{(1)}_L$~\cite{Kesden} which are typically discarded in CMB lensing power spectrum analyses. This provides a hint that the non-Gaussian signatures of CMB lensing are being considered as part of the SCALE signal. We also note that the SCALE procedure draws parallels with the estimator constructed in Ref.~\cite{Smith:2016lnt}, wherein the locally measured small-scale $(\ell \gtrsim 3000)$ temperature power varies across the sky due to the patchiness of the kinetic Sunyaev-Zel'dovich effect. The main difference is that here we correlate the fluctuations in power between large and small scales, whereas the kSZ estimator of Ref.~\cite{Smith:2016lnt} studies the angular power spectrum of the locally measured small-scale temperature power.

\section{Simulated observables}\label{sec:Sims}
We test SCALE on simulated CMB maps to determine the robustness of the method. The input power spectra for the simulated maps (shown in Figure~\ref{fig:InputPS}) in all of our analyses were generated with \texttt{CAMB}\footnote{\url{https://camb.info/}}~\cite{Lewis:1999bs,Howlett:2012mh} and the parameters listed in Table~\ref{tab:CAMBParams}. We choose parameters to approximately match results from the \planck\ results~\cite{Planck:2018nkj}, as well as accuracy factors suggested by~\cite{McCarthy:2022}. We generate all simulated raw CMB maps using the \texttt{rand\_map} method from \texttt{pixell}\footnote{\url{https://github.com/simonsobs/pixell}} at a resolution of 0.5'. Simulated maps are generally $10\degree \times 10\degree$ and centered at the equator. These smaller maps are well within the flat-sky approximation and can be quickly simulated in large quantities. The \texttt{rand\_map} method imposes repeating boundary conditions in each realization, so the filtering steps do not generate any edge effects. When gathering power spectrum and/or cross-spectrum statistics, we choose bins of $\Delta\ell$, $\Delta L$ and $\Delta\Lcheck$ that are integer multiples of the fundamental mode $\ell_\mathrm{fun}=36$ for our maps. This is to ensure that we gather values at bin widths which are commensurate with both the grids in which the realizations themselves were generated and in which the correlation statistics are evaluated. We apply lensing to the raw CMB maps using \texttt{pixell}'s \texttt{lensing} package, and a lens potential field corresponding to the $C_L^{\kappa\kappa}$ spectrum shown in Figure~\ref{fig:InputPS}. 

For every CMB map, we generate noise realizations with experiment-relevant values listed in Table~\ref{tab:Noise}. Configuration A represents a Stage III-like survey like ACT and SPT, while Configuration B is illustrative of the Simons Observatory \citep[SO,][]{SimonsObservatory:2018koc}. Configuration C gives noise and beam corresponding to CMB Stage IV-like properties \cite{CMB-S4:2016ple}, and Configuration D shows a slightly more futuristic experiment corresponding with some tests made for the gradient inversion estimator in \cite{Hadzhiyska:2019cle}. Configuration E represents a low-noise, high-resolution experiment like the proposed CMB-HD \cite{Sehgal:2019ewc}. We also briefly consider tests in the noise-free limit. 

We consider a range of current to future experiments, and we present a particular focus of results for Configuration D. We generally choose a window function for $W_\varsigma(\ell)$ to include modes $\ell_1 \in [6000, 8000]$ (unless otherwise shown) for a balance between being in a regime where the lensing signal is expected to be high, and noise is not too dominant (refer to Figure~\ref{fig:InputPS}). This reasoning is demonstrated against several choices of $\ell_1$ windows in Sec.~\ref{sec:results}, but we expect the SCALE methodology to be effective as long as $\ell_1$ satisfies our small-scale approximations (i.e., $\ell_1 \gg 2000$). We also generally choose $W_\lambda(\ell)$ to include modes $\ell_2 \in [0, 3000]$ to ensure that $\lambda$ maps include most of the information about the large-scale temperature gradient power.

\begin{table}
    \centering
    \begin{tabular}{|l|c|}
        \hline
        \rowcolor[HTML]{E7F9FF} Parameter & Value \\
        \hline
        \texttt{H0} & 67.5 km/s\\
        \rowcolor[HTML]{EFEFEF} \texttt{ombh2} & 0.022 \\
        \texttt{omch2} & 0.122 \\
        \rowcolor[HTML]{EFEFEF} \texttt{tau} & 0.06 \\
        \texttt{As} & 2e-9 \\
        \rowcolor[HTML]{EFEFEF} \texttt{ns} & 0.965 \\
        \texttt{r} & 0 \\
        \rowcolor[HTML]{EFEFEF} \texttt{lmax} & 20000 \\
        \texttt{lens\_potential\_accuracy} & 8 \\
        \hline
    \end{tabular}
    \caption{The set of non-default arguments given to \texttt{CAMB} when simulating power spectra chosen to approximately match results from~\cite{Planck:2018nkj}. Lensing accuracy parameters were chosen as suggested by~\cite{McCarthy:2022}.
    \label{tab:CAMBParams}}
\end{table}
%%%%%%%%%%%%%%%%%%%%%%%%

\begin{table}
    \centering
    \begin{tabular}{|c|c|c|l|}
        \hline
        \rowcolor[HTML]{E7F9FF} Config. & $ w~[\mu\mathrm{K}\mhyphen\mathrm{arcmin}] $ & $ b~[\mathrm{arcmin}] $ & Analogous Exp. \\
        \hline
        A & 10.5 & 1.3 & ACT \cite{ACT:2020gnv} \\
        \rowcolor[HTML]{EFEFEF} B & 6.3 & 1.4 & SO \cite{SimonsObservatory:2018koc} \\
        C & 1.5 & 1.4 & CMB-S4 \cite{CMB-S4:2016ple} \\
         \rowcolor[HTML]{EFEFEF} D & 1.0 & 1.0 & Comparison with \cite{Hadzhiyska:2019cle} \\
        E & 0.5 & 0.25 & CMB-HD \cite{Sehgal:2019ewc} \\
        \hline
    \end{tabular}
    \caption{The set of simulated noise configurations chosen to be representative of existing or upcoming experiments from ACT \citep[Configuration A,][]{ACT:2020gnv} to CMB-HD \citep[Configuration E,][]{Sehgal:2019ewc}
    \label{tab:Noise}}
\end{table}
%%%%%%%%%%%%%%%%%%%%%%%%%%%%%

We later compare the results of our SCALE lensing method to those of the Hu, DeDeo \& Vale \citep[HDV,][]{Hu:2007bt} quadratic estimator with the $TT$ fields. We choose to compare with a quadratic estimator since it provides well-understood and established benchmark. We choose the HDV quadratic estimator in particular due to its behavior in the small-scale regime. Small angular scale $(\ell \gtrsim 2000)$ contributions to the gradient power are removed in the HDV implementation to avoid a bias introduced by higher order cross-terms between the temperature gradient and the lensing convergence in this regime; this is less of a concern with the original Hu \& Okamoto estimators applied at larger angular scales (refer to Figure~\ref{fig:InputPS} and \cite{Hu:2007bt}). The HDV quadratic estimator and its principles have also been applied in studies of cluster lensing including some using \planck~\cite{Raghunathan:2018}, SPT data~\cite{DES:2017fyz}, ACT data~\cite{ACT:2020izl}, as well as forecasts of lensing results with the proposed CMB-HD experiment~\cite{Han:2022}. We perform reconstructions of the lensing convergence field $\hat{\kappa}$ with the HDV method for a subset of simulated lensed CMB maps, along with computations of the optimal noise ${N}_L^{(0)}$ and realization dependent noise $\hat{{N}}_L^{(0)}$ using the \texttt{symlens}\footnote{\url{https://github.com/simonsobs/symlens}} package. In particular, we choose \texttt{xmask} with $\ell_\mathrm{min}=2$ and $\ell_\mathrm{max}=3000$, \texttt{ymask} with $\ell_\mathrm{min}=2$ and $\ell_\mathrm{max}=10000$, and \texttt{kmask} with $L_\mathrm{min}=100$ and $L_\mathrm{max}=10000$. The first two masks are applied to each version of the temperature field of the $TT$ quadratic estimator, and the final mask is applied to the reconstructed convergence field. We apply our method to \num{100000} realizations for each suite of tests to obtain stable statistics of the SCALE output. In particular, we found that we need at least \num{100000} simulations to reach a converged inverse covariance matrix that we use later to compute signal-to-noise. We apply the HDV quadratic estimator on a subset containing \num{10000} of the full set of realizations when making comparisons, choosing a smaller sample size because it is computationally more intensive to run quadratic estimators. We also found that the inverse covariance matrix for the HDV output converges with a sample size of \num{10000}. Each set of simulations applying the SCALE procedure shares the following general flow:
\begin{enumerate}
    \item Generate primordial CMB temperature power spectrum $C_\ell^{TT}$, lensing power spectrum $C_L^{\phi\phi}$, and lensed CMB temperature power spectrum $\tilde{C}_\ell^{TT}$ with \texttt{CAMB}.
    \item Generate $N_\ell^{TT}$ according to one of the experiment configurations in Table~\ref{tab:Noise}.
    \item Compute $A_\Lcheck$ and $\expect{\Psi_\Lcheck}$ with the above power spectra. $A_\Lcheck$ are different for lensed/unlensed maps, and $\expect{\Psi_\Lcheck}=0$ for unlensed maps.
    \item For each of \num{100000} sims:
    \begin{enumerate}
        \item Generate realization of CMB temperature $T$ with $C_\ell^{TT}$, and lensing field $\phi$ with $C_L^{\phi\phi}$.
        \item Apply lensing to the CMB temperature field to get the lensed temperature field $\tilde{T}$. (Not done in the null test.)
        \item Generate noise field $N$ with $N_\ell^{TT}$, and add to $\tilde{T}$. (Add to $T$ in the null test with no lensing.)
        \item Follow the steps in Figure~\ref{fig:SCALESchematic} to estimate $\hat{\Psi}_\Lcheck$ for this given realization.
    \end{enumerate}
    \item[] The next three steps are unnecessary for SCALE, but are performed for \num{10000} iterations if we wish to compare SCALE with the HDV quadratic estimator.
    \begin{enumerate}[resume]
        \item Reconstruct the lensing convergence field $\hat{\kappa}$ with the HDV quadratic estimator described above.
        \item Estimate the lensing power spectrum $\hat{C}_L^{\kappa\kappa}$ with the reconstructed $\hat{\kappa}$ field.
        \item Compute the realization-dependent reconstruction noise $\hat{N}_L^{(0)}$.
    \end{enumerate}
\end{enumerate}
%
%%%%%%%%% Notation Table %%%%%%%%%%%%%%%%%%
\begin{table}
    \centering
    \begin{tabular}{|c|l|}
        \hline
        \rowcolor[HTML]{E7F9FF} Symbol & Description \\
        \hline
        $T$ & CMB temperature field \\
        \rowcolor[HTML]{EFEFEF} $\Tilde{T}$ & Lensed CMB temperature field \\
        $T_L$ & Large-scale temperature field \\
        \rowcolor[HTML]{EFEFEF} $T_S$ & Small-scale temperature field \\
        $\lambda$ & Large-scale temperature gradient power field \\
        \rowcolor[HTML]{EFEFEF} $\varsigma$ & Small-scale temperature gradient power field \\
        $\phi$ & CMB lensing potential field \\
        \rowcolor[HTML]{EFEFEF} $\kappa$ & CMB lensing convergence field \\
        $\ell$ & CMB multipole \\
        \rowcolor[HTML]{EFEFEF} $L$ & Lensing field multipole \\
        $\Lcheck$ & SCALE cross spectrum multipole \\
        \rowcolor[HTML]{EFEFEF} $\ell_1$ & Small-scale filter multipole \\
        $\ell_2$ & Large-scale filter multipole \\
        \rowcolor[HTML]{EFEFEF} $\nhat$ & Line-of-sight direction \\
        $C_\ell^{XY}$ & Cross (or auto) spectrum of fields $X$ and $Y$ \\
        \rowcolor[HTML]{EFEFEF} $N_\ell^{XX}$ & Noise spectrum of $XX$ \\
        $\Psi_\Lcheck$ & Normalized SCALE cross spectrum \\
        \rowcolor[HTML]{EFEFEF} $\Delta\Psi_\Lcheck$ & Minimum expected SCALE uncertainty \\
        $A_\Lcheck$ & SCALE normalization \\
        \rowcolor[HTML]{EFEFEF} $\mathbf{X}$ & Vector quantity $X$ \\
        $\grad X$ & Gradient of field $X$ \\
        \rowcolor[HTML]{EFEFEF} $\overline{X}$ & Average of quantity $X$ (possibly around $\nhat$) \\
        $\Delta X$ & Size/width of bin for quantity $X$ \\
        \rowcolor[HTML]{EFEFEF} $\hat{X}$ & Estimated quantity or reconstructed field $X$ \\
        $\expect{X}$ & Expected value of quantity $X$ \\
        \hline
    \end{tabular}
    \caption{Summary of notation relevant to the SCALE method for quick reference. Symbols appearing first take precedence in the case of apparent conflict.
    \label{tab:Notation}}
\end{table}
%%%%%%%%%%%%%%%%%%%%%%%%%%%%%

We provide a summary of all our SCALE-relevant notation in Table~\ref{tab:Notation} for quick reference. The summary statistics for SCALE $\hat{\Psi}_\Lcheck$ and the HDV quadratic estimator $\hat{C}_L^{\kappa\kappa}-\hat{N}_L^{(0)}$ over \num{100000} simulations are collected and presented in the following section. We make our code publicly available\footnote{\url{https://github.com/victorcchan/cmbpix}} along with example scripts and an example tutorial notebook.

\section{Results\label{sec:results}}

\begin{figure}
    \centering
    \includegraphics[width=1.0\columnwidth]{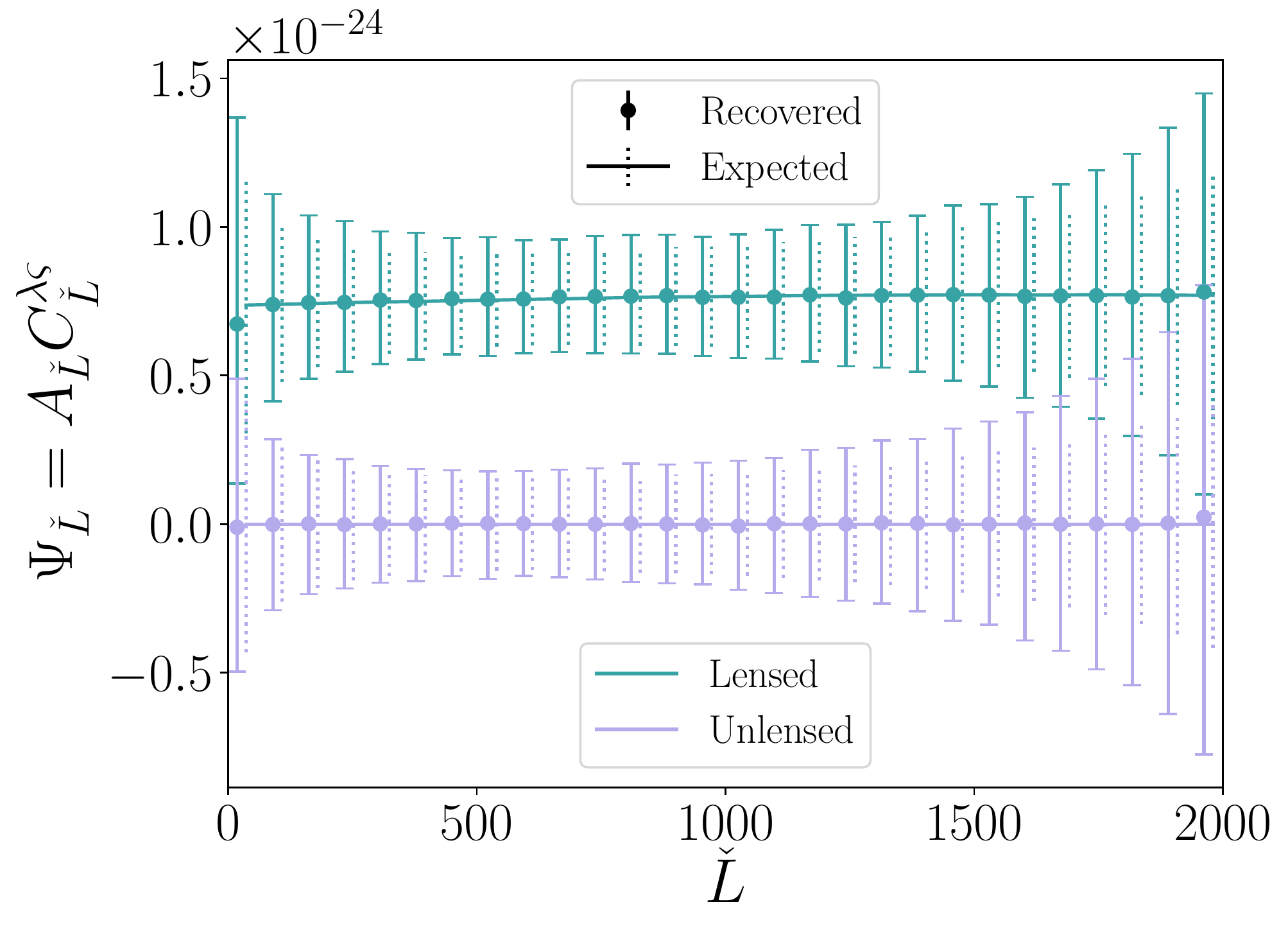}
    \caption{Comparison of expected and recovered $\Psi_\Lcheck$ band-powers with $\Delta\Lcheck=72$ from \num{100000} simulations of 100 sq.deg. temperature maps in noise configuration D. Estimates of recovered $\hat{\Psi}_\Lcheck$ are the median and 68\% scatter of the band-power at each bin.}
    \label{fig:PsiL}
\end{figure}

The SCALE estimator is exceptional at detecting the presence of small-scale lensing in CMB temperature maps. We illustrate this in Figure~\ref{fig:PsiL}, which shows the summary statistics of recovered $\hat{\Psi}_\Lcheck$ from \num{100000} simulations of $10\degree \times 10\degree$ lensed and unlensed CMB temperature maps in the experiment D noise configuration. The combination of all bins in Figure~\ref{fig:PsiL} corresponds to a detection of lensing in 100 sq.deg. maps over the null test with an approximate SNR of 9. We also show the expected theoretical $\expect{\Psi_\Lcheck}$ computed with Eq.~\eqref{eq:PsiExpected1}. The vertical extent of the error bars represents the 68\% width of the distribution (centered at the median within the $\Lcheck$ bin) of estimated $\hat{\Psi}_\Lcheck$ band-powers, and they describe the statistical scatter of estimated band-powers for a given CMB realization of similar total area. The scatter of these band powers is comparable to, but slightly in excess of, the minimal expectation $\Delta\Psi_\Lcheck$ given by Eq.~\eqref{eq:PsiLError}. Even with relatively small maps of the CMB temperature, the SCALE estimator is able to make a clear distinction of whether or not lensing is present in maps generated with the parameters described in Section~\ref{sec:Sims} and Figure~\ref{fig:InputPS}.

%%%%%%%%%%%%%%%%% Psi_L %%%%%%%%%%%%%%%%%%
%%%%%%%%%%%%%%%%%%%%%%%

%%%%%%%%%%%%%%%%%%%%%%%%%%% Shifting l1 windows %%%%%%%%%%%%%%%%%%%%%
\begin{figure}
    \centering
    \includegraphics[width=\columnwidth]{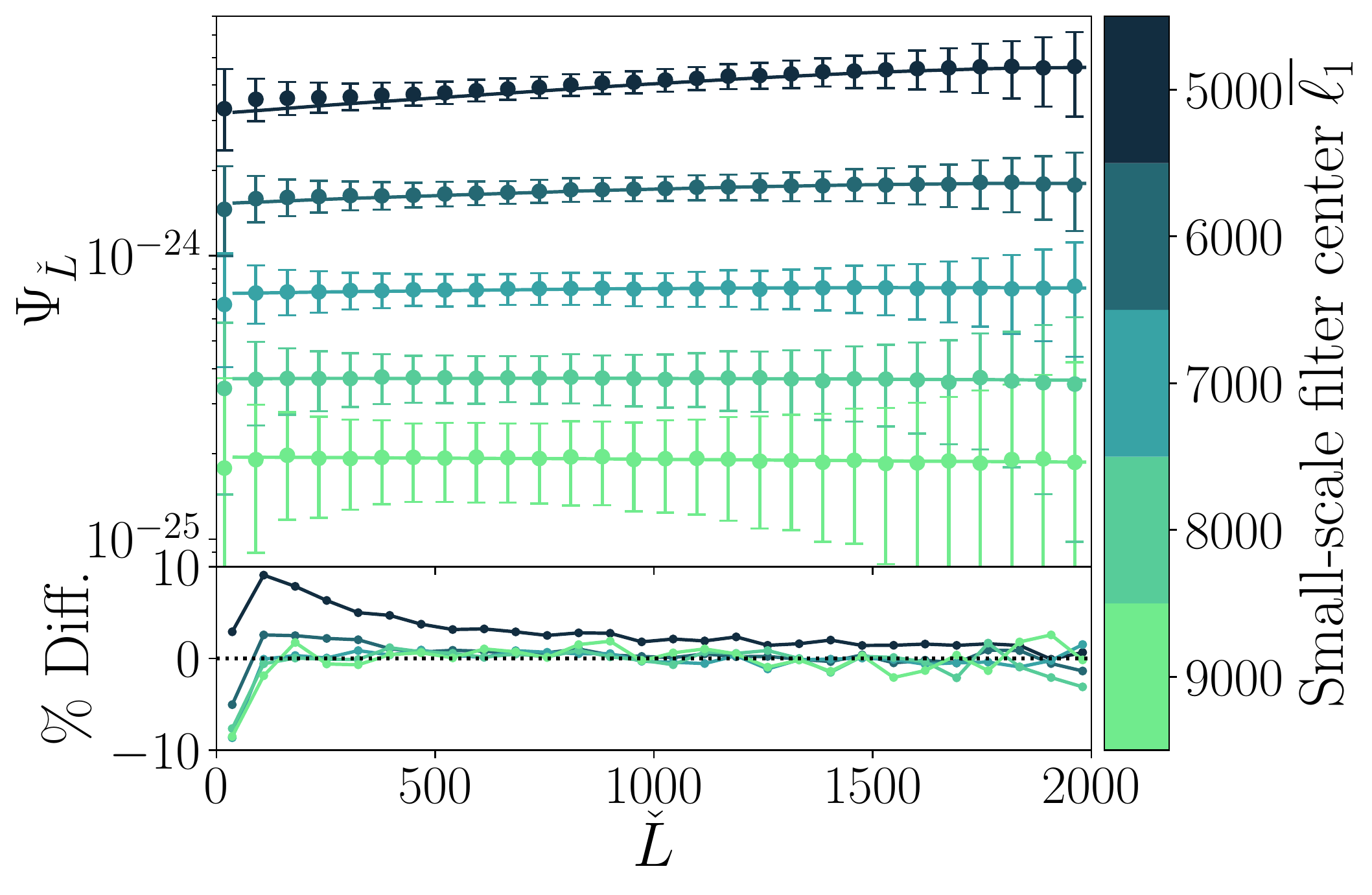}
    \caption{\textit{Top}: Comparisons of expected $\Psi_\Lcheck$ and spread from theory (solid lines with dotted error-bars) and median recovered $\Psi_\Lcheck$ and 68\% spread from simulation (points with capped error-bars) when shifting the centre of the $\ell_1$ window which defines the small-scale filter while keeping the width of the filter constant at $\Delta\ell_1=2000$. Error-bars shown are reduced by a factor of 2 for improved visual comparison. \textit{Bottom}: The bias of the recovered $\Psi_\Lcheck$ when compared the the expected $\Psi_\Lcheck$ shown as a percentage of the total signal. At fixed window width, increasing the central $\ell_1$ reduces the strength of the recovered signal. }
    \label{fig:l1Shift}
\end{figure}
%%%%%%%%%%%%%%%%%%%%%%%%%%%%%%%%%%%%%%%%%

%%%%%%%%%%%%%%%%%%%%%%%%%%% Stretching l1 windows %%%%%%%%%%%%%%%%%%%%%
\begin{figure}
    \centering
    \includegraphics[width=\columnwidth]{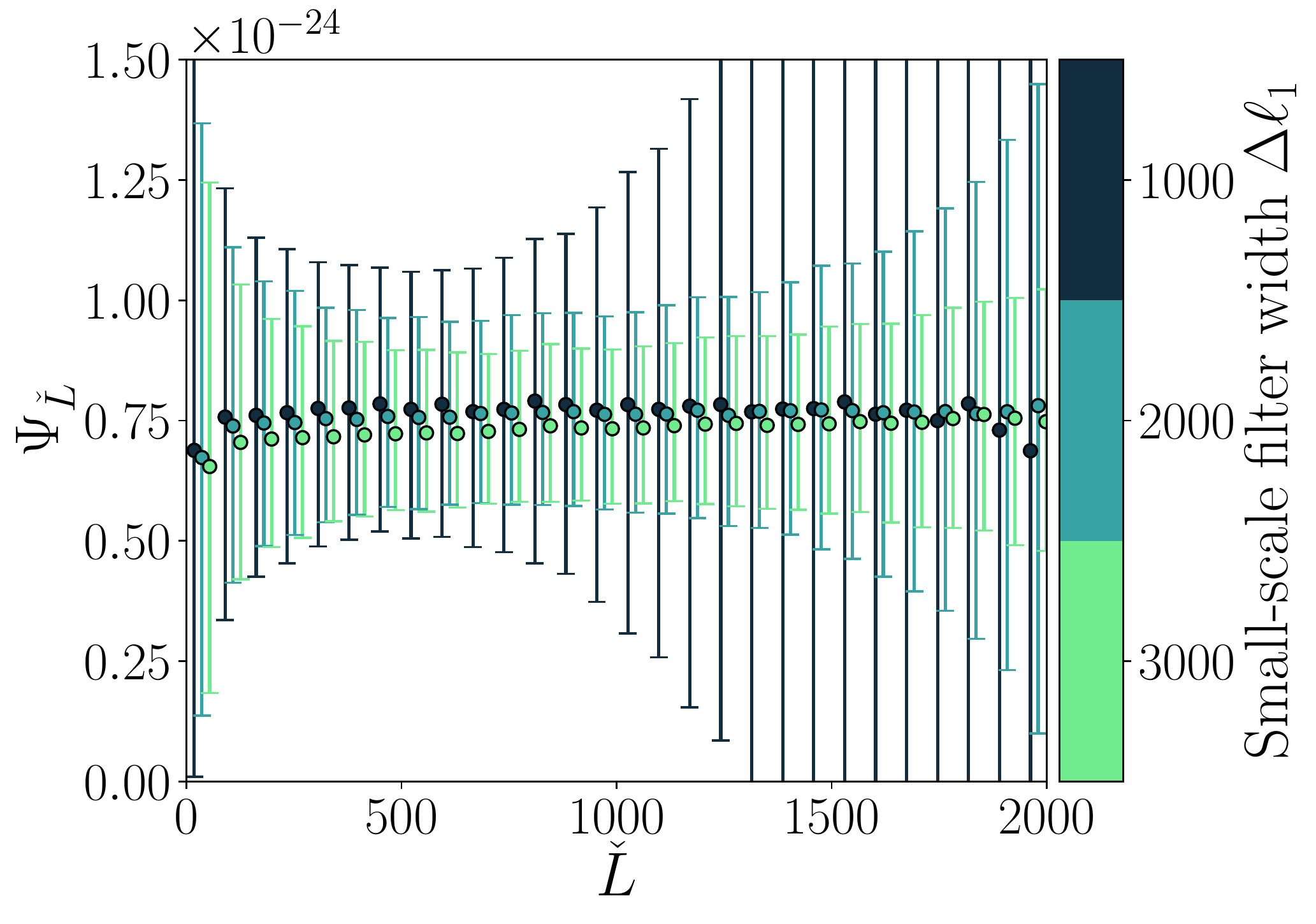}
    \caption{As a Complement to Figure~\ref{fig:l1Shift}, the comparisons of expected $\Psi_\Lcheck$ and spread from theory (solid lines with dotted error-bars) and median recovered $\Psi_\Lcheck$ and 68\% spread from simulation (points with capped error-bars) when altering the size of the small-scale filter window  $\Delta\ell_1$ used to compute the signal, this time keeping the centre of the $\ell_1$ window at $\overline{\ell_1}=7000$. For a fixed central $\ell_1$, reducing the width of the $\ell-$window merely increases the error bar on the recovered signal. }
    \label{fig:l1Stretch}
\end{figure}
%%%%%%%%%%%%%%%%%%%%%%%%%%%%%%%%%%%%%%%%%

In principle, the SCALE method can be applied to any $\ell_1$ regime so long as the small-scale lensing approximations are appropriate. Figure~\ref{fig:l1Shift} shows comparisons between expected $\expect{\Psi_\Lcheck}$ computed with Eq.~\eqref{eq:PsiExpected1} and Eq.~\eqref{eq:PsiLError} and recovered $\hat{\Psi}_\Lcheck$ band-power statistics from simulations for different shifts in the small-scale $\ell_1$ window while keeping $\Delta\ell_1=2000$. This roughly corresponds to shifting (in the same direction) which $C_\ell^{TT}$, $N_\ell^{TT}$ and $C_L^{\kappa\kappa}$ modes contribute to $\Psi_\Lcheck$. The expected and recovered band-powers agree to the same extent as the results from Figure~\ref{fig:PsiL}. The recovered band-powers begin to exhibit a positive bias as the $\ell_1$ window is shifted towards lower $\ell$, which can be explained by a departure from the small-scale lensing approximations made in Section~\ref{sec:method/intro}. In particular, the Taylor series expansion in Eq.~\eqref{eq:TaylorExpansion} becomes inaccurate at scales $\ell \sim 2000$ because of the similarity of scales with the average deflection angle \cite{Lewis:2006fu}. Higher accuracy in this regime would require consideration of higher-order terms of the expansion. The overall amplitude of each $\Psi_\Lcheck$ curve decreases as the $\ell_1$ window shifts to higher $\ell$, which reflects the shape of the $C_L^{\kappa\kappa}$ lensing power spectrum in Figure~\ref{fig:InputPS}. The statistical spread of $\Psi_\Lcheck$ band-powers grows as the $\ell_1$ window is shifted to higher $\ell$ because of increased contributions from experiment noise $N_\ell^{TT}$ at high $\ell$ (see Figure~\ref{fig:InputPS}). The default $6000 < \ell_1 < 8000$ window presented in Figure~\ref{fig:PsiL} offers a balance between satisfying the small-scale lensing approximations while not appearing to be statistically dominated by experiment noise. 

Mathematically, Eq.~\eqref{eq:ALDef1} and Eq.~\eqref{eq:PsiExpected1} are constructed such that the SCALE output is a normalized estimate of the average lensing power within and slightly around the small-scale $\ell_1$ window. The $\ell_1$ window width determines how many $C_L^{\kappa\kappa}$ modes contribute to what we consider signal, but the amplitude of $\Psi_\Lcheck$ centered at the same $\overline{\ell_1}$ should not change significantly after normalization with $A_\Lcheck$. Similarly, we add more contributions from $N_\ell^{TT}$ modes, but they are not expected to strongly correlate between large/small-scales. The overall effect of widening the $\ell_1$ window is to use the increased presence of lensing signal to reduce statistical scatter of recovered $\Psi_\Lcheck$ band-powers. Figure~\ref{fig:l1Stretch} illustrates this quite well by comparing $\Psi_\Lcheck$ after narrowing or widening $\Delta\ell_1$ while keeping the window centered at $\overline{\ell_1}=7000$. The overall amplitude of the normalized $\Psi_\Lcheck$ curves appears mostly unchanged, and a wider window does indeed result in tighter distributions of recovered $\Psi_\Lcheck$ band-powers. The changes we do see in the amplitude of $\Psi_\Lcheck$ are set by the shape of the underlying lensing power spectrum (i.e., the slope of $C_L^{\kappa\kappa}$ is slightly steeper on one side of the bin center when compared to the other).

%%%%%%%%%%%%%% Lensing Covariance Matrices %%%%%%%%%%%%%
\begin{figure}[h!]
    \centering
    \includegraphics[width=0.9\columnwidth]{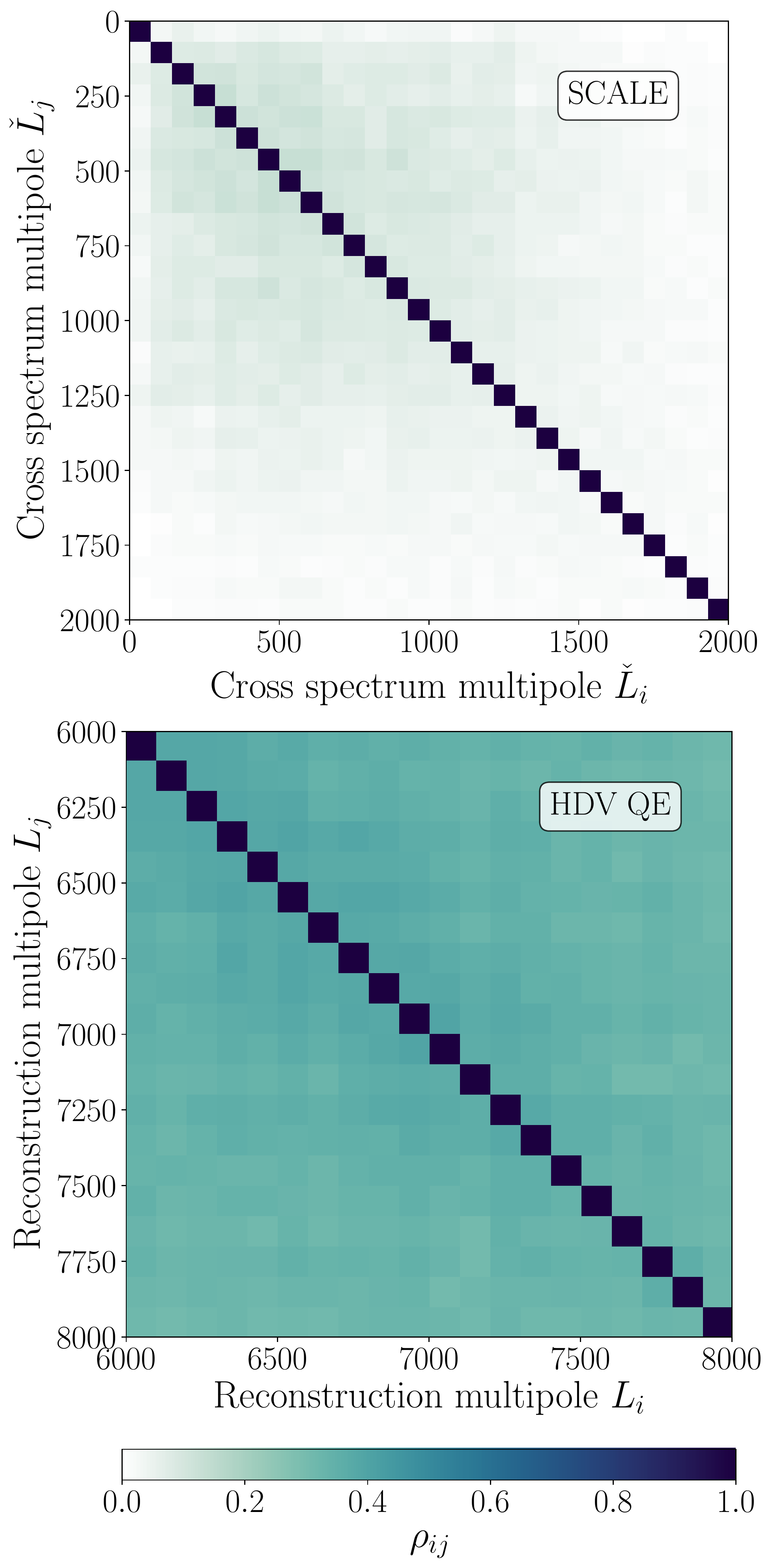}
    \caption{\textit{Top:} Correlation matrix of the SCALE estimator output $\Psi_\Lcheck$ over \num{100000} simulations in noise configuration D. The typical variance at a $\Lcheck$ bin of width $\Delta\Lcheck=72$ is of order $\sim(10^{-25})^2$. \textit{Bottom:} Correlation matrix of the HDV quadratic estimator reconstructed noise subtracted lensing potential power spectra 
    %$\hat{C}_L^{\phi\phi} - \mathcal{N}_L^{\phi\phi}$
    $\hat{C}_L^{\phi\phi} - \hat{N}_L^{(0),\phi\phi}$ for a subset containing \num{10000} of the above simulations. The typical variance at a $L$ bin of width $\Delta L=100$ is of order $\sim(10^{-25})^2$.}
    \label{fig:Covmats}
\end{figure}
%%%%%%%%%%%%%%%%%%%%%%%%%%%%

While the goal of conventional methods is to reconstruct the underlying lensing field, the SCALE method's output $\Psi_\Lcheck$ is an indirect estimate of the statistics of the lensing field. Figure~\ref{fig:Lcheck} illustrates the space of $\Lcheck$ modes in which the small-scale temperature power fluctuations correlate with those on the large-scales. These $\Lcheck$ modes are \textit{not equivalent} to the space of $L$ modes describing the lensing field itself. As shown in Eq.~\eqref{eq:PsiExpected1}, each band-power of $\Psi_\Lcheck$ contains information from a wide range of lensing field statistic modes $C_L^{\kappa\kappa}$. Figure~\ref{fig:Covmats} shows the covariance between estimated $\Delta\Lcheck=72$ band-powers using SCALE on \num{100000} simulations with noise configuration D, as well as the covariance between estimated $\Delta L=100$ band-powers of noise-subtracted lensing power spectra $\hat{C}_L^{\phi\phi} - \hat{N}_L^{(0),\phi\phi}$ from \num{10000} reconstructions using the HDV quadratic estimator on a subset of simulations. 

Each set of recovered $\hat{\Psi}_\Lcheck$ band-powers appears to contain correlations induced by lensing that have weak correlations between low $\Lcheck$ modes, and those at higher $\Lcheck$ modes appear mostly independent from other modes. This is in contrast to the estimated band-powers of the quadratic estimator, which have relatively strong correlations between all bands.

%%%%%%%%%%%% Lensing Signal to Noise %%%%%%%%%%%%%%
\begin{figure}
    \centering
    \includegraphics[width=0.9\columnwidth]{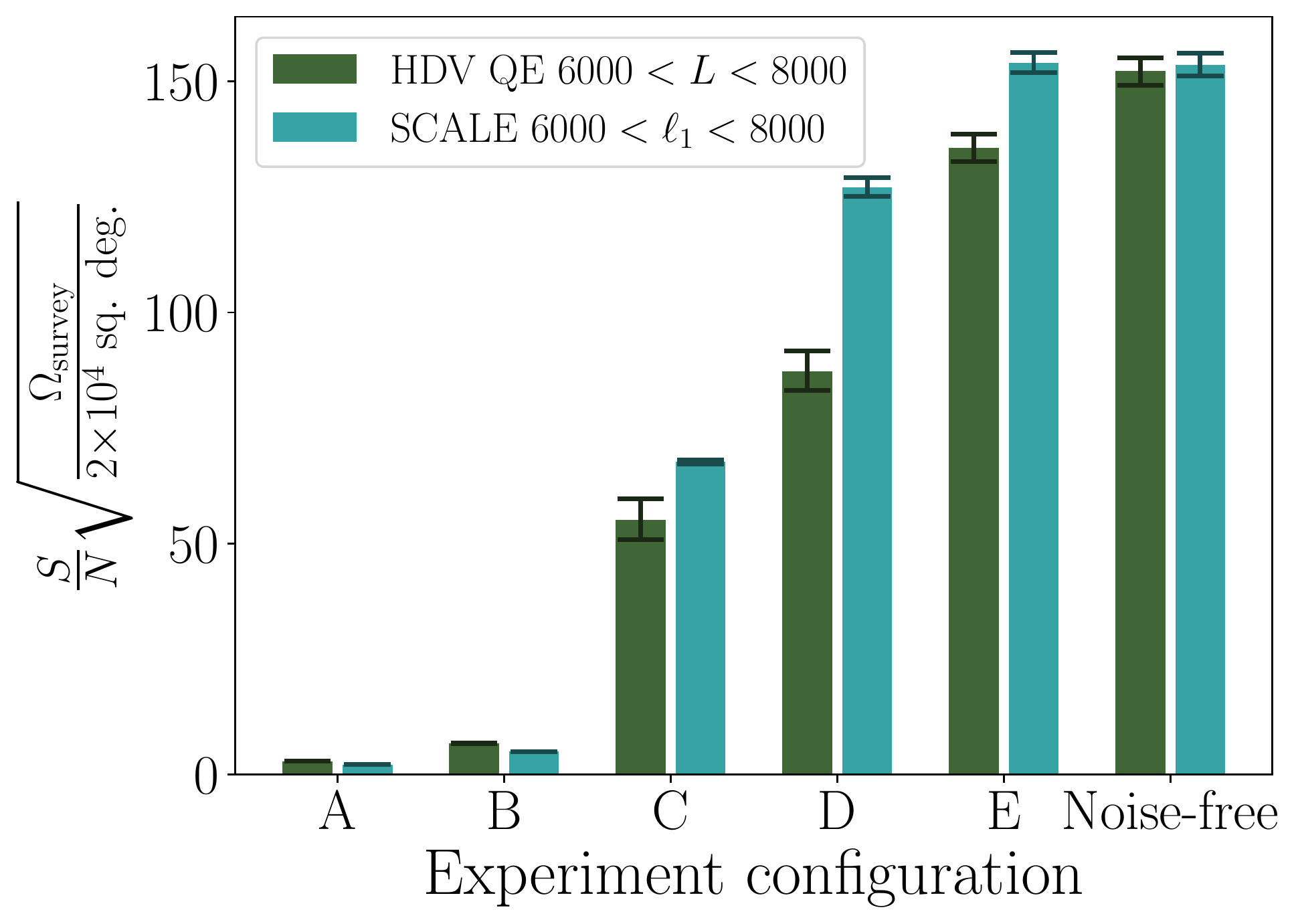}
    \caption{Comparison of the values of the signal-to-noise ratio (SNR, Eq.~\eqref{eq:SNR}) across noise configurations between the SCALE estimator for a small-scale window $W_\varsigma(6000 < \ell < 8000)$ and the HDV quadratic estimator applied to $6000 < L < 8000$. SCALE bars indicate the median and 68\% range of the bootstrap distribution for SNRs computed \num{1000} times with a set of \num{100000} simulations. HDV bars indicate the median and 68\% range of the bootstrap distribution for SNRs computed \num{1000} times with a set of \num{10000} simulations. Each realization has a map area of 100 sq.deg., and SNR values are scaled up to \num{20000} sq.deg.
    }
    \label{fig:SNRcomparison}
\end{figure}
%%%%%%%%%%%%%%%%%%%%%%%%%%%%%%%

%%%%%%%%% SNR Table %%%%%%%%%%%%%%%%%%
\begin{table}
    \centering
    \begin{tabular}{|l|c|c|}
        \hline
        \rowcolor[HTML]{E7F9FF} Config. & HDV QE $S/N$ & SCALE $S/N$ \\
        \hline
        A & 2.9 & 2.1 \\
        \rowcolor[HTML]{EFEFEF} B & 6.8 & 5.0 \\
        C & $55.1^{+4.7}_{-4.2}$ & $67.6^{+0.4}_{-0.4}$ \\
         \rowcolor[HTML]{EFEFEF} D & $87.3^{+4.5}_{-4.2}$ & $127.0^{+2.1}_{-1.9}$ \\
        E & $135.7^{+3.0}_{-3.0}$ & $154.0^{+2.2}_{-2.0}$ \\
        \rowcolor[HTML]{EFEFEF} Noise-free & $152.2^{+2.9}_{-3.1}$ & $153.6^{+2.4}_{-2.5}$ \\
        \hline
    \end{tabular}
    \caption{Computed signal-to-noise ratio (SNR, Eq.~\eqref{eq:SNR}) across noise configurations between the SCALE estimator for a small-scale window $W_\varsigma(6000 < \ell < 8000)$ and the HDV quadratic estimator applied to $6000 < L < 8000$. Values are the median and 68\% range of the bootstrap distribution for SNRs computed \num{1000} times with a set of \num{100000} simulations for SCALE and \num{10000} simulations for the HDV QE. No uncertainty is shown if the 68\% range of the bootstrap distribution is smaller than the significant figures provided. Each realization has a map area of 100 sq.deg., and SNR values are scaled up to \num{20000} sq.deg.
    \label{tab:SNR}}
\end{table}
%%%%%%%%%%%%%%%%%%%%%%%%%%%%%

Each covariance matrix $\mathbf{C}$ allows us to compute a signal-to-noise ratio 
\begin{equation}
    \mathrm{SNR} = \sqrt{a^\mathrm{T}\mathbf{C}^{-1}a}
    \label{eq:SNR}
\end{equation}
using either $a=\expect{\Psi_\Lcheck}$ for SCALE or $a=C_L^{\phi\phi}$ for the quadratic estimator. Using covariance matrices for SCALE and HDV QE outputs of \num{100000} and \num{10000} simulations respectively with the noise configurations in Table~\ref{tab:Noise}, we compute the expected signal-to-noise ratios and compare them in Figure~\ref{fig:SNRcomparison} as well as Table~\ref{tab:SNR}. We scale the covariance matrices used to compute each of these signal-to-noise values by a factor of map-area ratios from 100 sq.deg. to \num{20000} sq.deg.

The results from the quadratic estimator directly retrieve information from modes of the lensing field between $6000 < L < 8000$ along with some covariance from modes outside of this band; however, we note that we have neglected to consider the effect of subtracting the higher order $N_L^{(1)}$ bias in this simple calculation as is typically done in QE analyses. The same modes of the lensing field are the main contribution to the SCALE results with the $6000 < \ell_1 < 8000$ window due to our choice of $W_\varsigma(\ell)$, but the set of $\Psi_\Lcheck$s includes contributions from modes of the lensing field outside of this band as shown in Eq.~\eqref{eq:PsiExpected1}. 

The reverse is also true: lensing modes within $6000 < L < 8000$ would also contribute to a lesser extent in other implementations of SCALE with a different choice in $\ell_1$ range. These modes would make up a small part of the SCALE signal if we choose instead to filter for $W_\varsigma(8000 < \ell < 10000)$ rather than the filter choice we make in this paper for $W_\varsigma(6000 < \ell < 8000)$. These contributions to the SCALE estimator from modes outside the filtered band make a direct comparison between both SCALE and other approaches difficult, as it is non-trivial to restrict SCALE to include information only from a certain subset of $L$ modes from the lensing field.

\begin{figure}
    \centering
    \includegraphics[width=0.9\columnwidth]{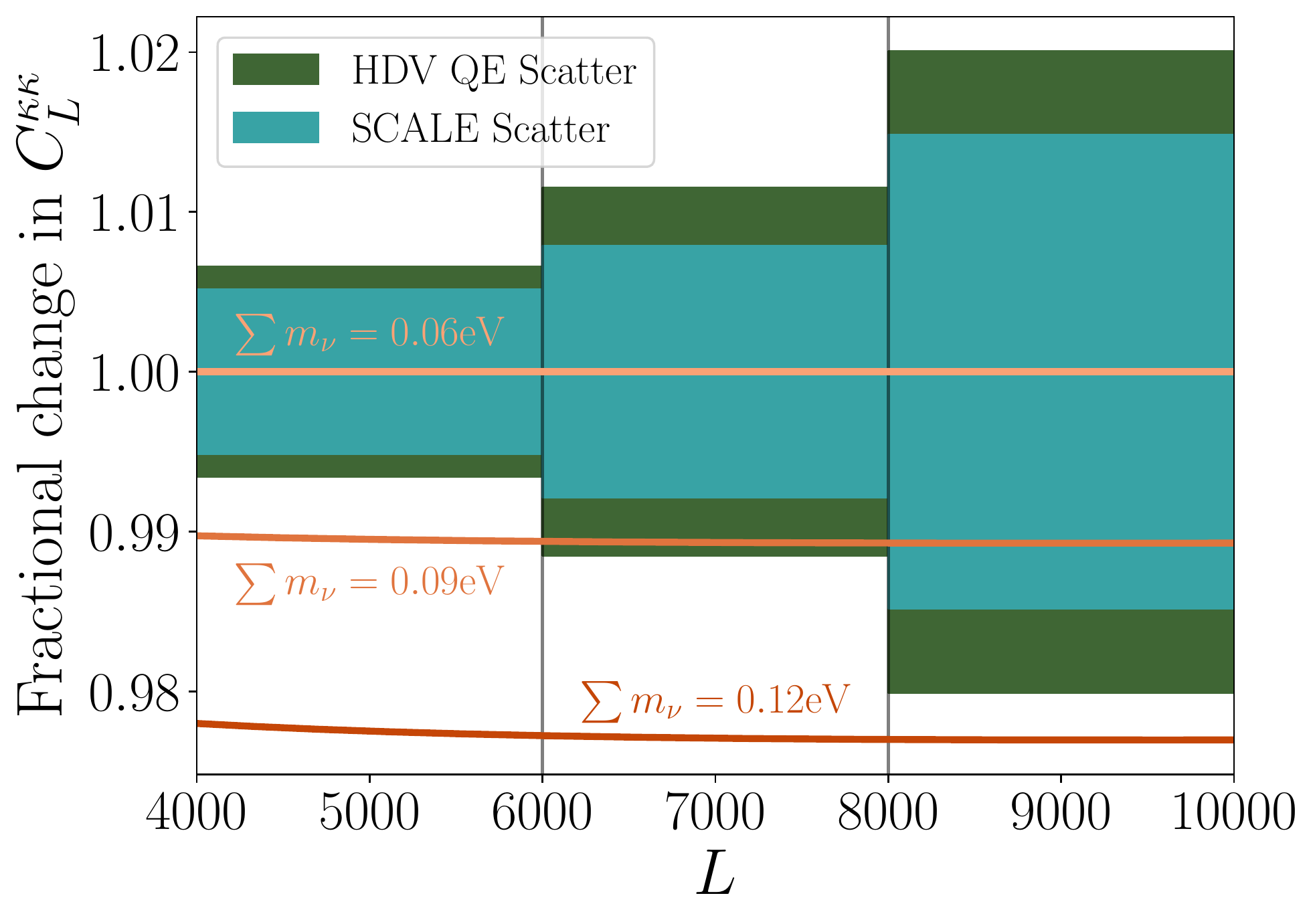}
    \caption{Comparison of SCALE vs HDV QE fractional error bands alongside fractional changes in the lensing convergence power $C_L^{\kappa\kappa}$ with neutrino mass. Note that $\sum m_\nu=0.06~\mathrm{eV}$ is the fiducial model used in our previous analyses.}
    \label{fig:CLkkerrorbar}
\end{figure}

Finally, we consider in a simple example the SCALE method's ability to discriminate between cosmological models that predict changes in the shape/amplitude of the matter power spectrum $P(k)$, and by extension, the lensing power $C_L^{\phi\phi}$ or $C_L^{\kappa\kappa}$ by adjusting the total sum of neutrino masses $\sum m_\nu$. A higher neutrino mass produces effects similar to warm or fuzzy dark matter, suppressing structure formation at small scales, but one key feature is that the lensing power $C_L^{\kappa\kappa}$ is suppressed similarly at high $L$. Figure~\ref{fig:CLkkerrorbar} compares a couple of models with neutrino mass heavier than our fiducial model. We see that the fractional changes in $C_L^{\kappa\kappa}$ do not contain much shape information, but in principle, different choices small-scale windows can elucidate potential shape information. We place \textit{approximate} fractional error bands of $\Delta L = 2000$ for SCALE and the HDV QE on the assumption that the SNR values for noise configuration D can be taken at face value. In other words, each fractional error band shown in Figure~\ref{fig:CLkkerrorbar} is calculated as $1 / \mathrm{SNR}$, centered on the fiducial curve.

While Figure~\ref{fig:CLkkerrorbar} is not meant to be a forecast of either SCALE or QE performance, it provides some insight into the distinguishing power of each method. We note that the SCALE estimator exists in a separate space of $\Lcheck$ modes that are each an estimate of a weighted sum of $C_L^{\phi\phi}$ (and by extension $C_L^{\kappa\kappa}$) as prescribed by Eq.~\eqref{eq:PsiExpected1}. We leave full parameter constraints and a more thorough comparison between methods for future work.

%%%%%%%%%%%%%%%%%%%%%%% Error on Lensing Spectrum %%%%%%%%%%%%
%%%%%%%%%%%%%%%%%%%%%%%%%

\section{Discussion \& Conclusion \label{sec:discuss}}

In this paper, we:
\begin{itemize}

    \item showed that fluctuations in the local small-scale $(\ell \gg 3000)$ CMB temperature power are intricately tied to the variations in the local large-scale temperature gradient through correlations induced by lensing.

    \item confirmed that this correlation is readily detectable in the pixel-space statistics of a lensed CMB temperature map, and there is no discernable correlation in a CMB temperature map without lensing (see Figure~\ref{fig:ScatterPlotPatches}).

    \item visualized correlations between a cross spectrum of large/small-scale CMB temperature power in Figure~\ref{fig:Lcheck}.

    \item created the Small Correlated Against Large Estimator (SCALE) which efficiently applies various filters to pick out the relevant small/large-scale in a CMB temperature map, and computes their cross spectrum (Figure~\ref{fig:SCALESchematic}).

    \item demonstrated that the SCALE method effectively recovers the expected statistics of underlying lensing fields, which matches well with insignificant bias against analytic forms (Figure~\ref{fig:PsiL}).

    \item tested the properties of the SCALE estimator against different choices of filtering scales (Figures~\ref{fig:l1Shift}-\ref{fig:l1Stretch}).

    \item determined that the SCALE method can outperform (by a factor of up to 1.5 in signal-to-noise) estimates of the CMB lensing power spectrum $C_L^{\phi\phi}$ through reconstruction with quadratic estimators in noise configurations similar to future experiments.
    
\end{itemize}

We find that significant lensing signals at modes $L > 3000$ can be recovered by exploiting the dependence of the CMB temperature power at similar scales $\ell > 3000$ on fluctuations of the CMB temperature gradient (Eq.~\eqref{eq:TaylorExpansion}, illustrated by Figure~\ref{fig:Lcheck}). A key advantage of this method is the expectation that any noise, foregrounds, and other CMB secondaries present in observations of the CMB temperature are not expected to correlate with the CMB gradient fluctuations themselves. Foregrounds in particular are known to be a nuisance at all angular scales (e.g.,~\cite{vanEngelen:2013rla,Osborne:2013nna}), but there have been significant advancements in foreground cleaning techniques in tandem with multi-frequency observations from space experiments such as \planck. We expect foregrounds to be well controlled at angular scales relevant to the construction of the large-scale $\lambda$ maps~\cite{Madhavacheril:2018,Beck:2020,Abylkairov:2021,Darwish:2021}. We expect the effects of foregrounds and foreground cleaning to change the noise structure in the small-scale $\varsigma$ maps, but these contributions should not correlate with $\lambda$ if the foregrounds have been cleaned properly at large angular scales. At these small scales, telescope systematic effects such as the differential and boresight pointing become more important and need to be modeled correctly to remain below a level of 1$\sigma$ \cite{Mirmelstein:2020pfk}.

We first demonstrated the correlation between large/small-scales in the CMB temperature field with real-space statistics to detect the presence of lensing in a small temperature map covering 100 square degrees. The SCALE procedure builds on the intuitions of the real-space method from Section~\ref{sec:method/real}, and it can successfully quantify the correlations induced by lensing in line with expectations.

A simple comparison of signal-to-noise for configurations similar to present and future experiments reveals that the SCALE method demonstrates a marked improvement over the effectiveness of traditional quadratic estimators at low noise levels in the small-scale regime.
We do not expect that SCALE will serve as a replacement for existing lensing reconstruction techniques.  Map-level reconstruction is useful for delensing and cross-correlation studies, and there are existing reconstruction techniques that are optimal across a wide range of angular scales.  SCALE provides the most benefit in the small-scale and low noise lensing regime, so the most precise lensing measurements are likely to come from a combination of different techniques applied to different scales.  This could be achieved, for example, by utilizing an estimate of the lensing map from a quadratic estimator, using the estimated lensing map to delens the CMB temperature, and then applying SCALE to estimate the power spectrum of the low noise, small-scale lensing modes that remain in the delensed map. We wish to highlight the simplicity of the SCALE pipeline's steps, which allows it to be quickly applied to any given CMB temperature map. This is in contrast to the maximum likelihood and maximum a posteriori methods which have been shown to be optimal, but they are computationally expensive to perform. These methods, in addition to the Bayesian and Gradient Inversion methods, reconstruct the underlying lensing field $\phi$, which can then be cross-analyzed with other observations such as galaxy clusters. SCALE does not reconstruct a map of the lensing field, but it is a simple and fast method of effectively recovering the statistics of the underlying lensing field to levels of accuracy and precision beyond what is capable with QE techniques.

The SCALE method presents an optimistic outlook for the future of CMB lensing science, providing a fresh opportunity to make high-quality estimates of lensing statistics using a relatively straightforward procedure in a regime that has historically been limited in CMB-only techniques due to limits in techniques and observational noise. The small-scale regime is particularly exciting because the lensing statistics here are sensitive to a wide-range of dark matter and gravitational clustering phenomena.

\section*{Acknowledgments}
The authors would like to thank Kendrick Smith for early discussions that inspired this work. 
We would also like to thank Dongwon Han, Mat Madhavacheril, Neelima Sehgal, and Blake Sherwin for fruitful discussions. 
VCC and RH would like to thank Jo Bovy, Keith Vanderlinde and Marten van Kerkwijk for insightful feedback about this work.  VCC would like to thank Emily Deibert, Tom\'as Cassanelli, Taylor Kutra, Natalie Price-Jones, Thierry Serafin Nadeau, Jennifer Scora, Ariel Amaral, James Lane, Steffani Grondin, Samantha Berek, Ayush Pandhi, Dang Pham, Phil Van-Lane, Mairead Heiger, Anika Slizewski, Alexander Laroche, Alex Lagu\"e, Margaret Ikape, Martine Lokken, Harrison Winch, Simran Nerval, Jason Leung, Vismay Shah, Visal Sok, Bruce Wu, Keir Rogers, Yilun Guan, Fraser Evans, and Ziggy Pleunis for thoughtful discussions as well as emotional support during this study. VCC would also like to thank Heather Chan, Bonnie Chu, Marrick Chan, Alvin Chan, Clarence Fung, Noel Ng, Dan Nguyen, Lillian Yeung, Chanakya Chegowni, Hilton Law, Ryan Shen, Timothy Ho, Tony Cheng, and Kyle Ngo for providing emotional support throughout this project. VCC received support from the Ontario Graduate
Scholarship/Queen Elizabeth II/Walter John Helm Graduate Scholarships In Science And Technology. Canadian co-authors acknowledge support from the Natural Sciences and Engineering Research Council of Canada (NSERC).  RH is supported by Natural Sciences and Engineering Research Council of Canada Discovery Grant Program and the Connaught Fund.
JM is supported by the US~Department of Energy under Grant~\mbox{DE-SC0010129}.
Some computational resources for this research were provided by SMU’s Center for Research Computing.
The Dunlap Institute is funded through an endowment established by the David Dunlap family and the University of Toronto. 
The authors at the University of Toronto acknowledge that the land on which the University of Toronto is built is the traditional territory of the Haudenosaunee, and most recently, the territory of the Mississaugas of the New Credit First Nation. They are grateful to have the opportunity to work in the community, on this territory.
Computations were performed on the SciNet supercomputer at the SciNet HPC Consortium. SciNet is funded by: the Canada Foundation for Innovation; the Government of Ontario; Ontario Research Fund - Research Excellence; and the University of Toronto. 
\newpage

\appendix

%%%%%%%%%%%%%%%%
%%%%%%%%%%%%%%%%
\onecolumngrid
\section{Derivation of SCALE}
\label{sec:SCALEDerivation}
\label{sec:scalederivation}

We wish to construct an estimator of the small scale lensing power.
We will construct the estimator using the cross correlation between a field constructed from the square of small-scale temperature fluctuations and a field constructed from the square of large-scale temperature fluctuations.
In the very small-scale regime, where the effects of lensing dominate the temperature power spectrum, the temperature power is proportional to the product of the large-scale temperature gradient power and the small-scale lensing deflection power.
The motivation of the estimator that we construct here is that the small-scale lensing-induced temperature power is non-Gaussian; the locally measured small-scale temperature power is correlated with variations in the large-scale temperature gradient, and the relation between the two is proportional to the small-scale lensing power.

Let us first define a field $\varsigma$ defined by the locally measured small-scale temperature gradient power
\begin{align}\label{eq:SSPowerDef}
    \varsigma(\vL) = \int \frac{\dd^2\vl_1}{2\pi} g(\vl_1, \vL) \left(\vl_1 \cdot (\vl_1-\vL) \right) T(\vl_1) T(\vL-\vl_1) \, ,
\end{align}
where $g$ is a filter applied to the small-scale temperature fluctuations, to be determined in what follows.
We will expand the small-scale temperature fluctuations to first order in the lensing gradient
\begin{align}
    \tilde{T}(\vl) = \int \frac{\dd^2\vl_2}{2\pi} \left(\vl_2 \cdot (\vl_2-\vl) \right) T(\vl_2) \phi(\vl-\vl_2) \, ,
\end{align}
where we have dropped the unlensed small-scale temperature, since it is assumed to be negligible compared to the lensing contribution on very small scales.
Inserting this into $\varsigma$ gives
\begin{align}
    \varsigma(\vL) =& \int \frac{\dd^2\vl_1}{2\pi} 
    g(\vl_1, \vL) \left(\vl_1 \cdot (\vl_1-\vL) \right)
    \int \frac{\dd^2\vl_2}{2\pi} 
    \int \frac{\dd^2\vl_3}{2\pi} 
    \left(\vl_2 \cdot (\vl_2-\vl_1) \right)
    \left(\vl_3 \cdot (\vl_3-\vL+\vl_1) \right) \nonumber \\
    & \times T(\vl_2) T(\vl_3) \phi(\vl_1-\vl_2) \phi(\vL-\vl_1-\vl_3) \, .
\end{align}
We will be interested in an estimate of the small-scale lensing power, rather than the realization of the lensing potential, so we will take an average over lensing realizations
\begin{align}
    \varsigma(\vL) =& \int \frac{\dd^2\vl_1}{2\pi} 
    g(\vl_1, \vL) \left(\vl_1 \cdot (\vl_1-\vL) \right) \nonumber \\
    & \times \int \frac{\dd^2\vl_2}{(2\pi)^2} 
    \left(\vl_2 \cdot (\vl_2-\vl_1) \right)
    \left( (\vL-\vl_2) \cdot (\vl_1-\vl_2) \right) 
    T(\vl_2) T(\vL-\vl_2) C_{|\vl_1-\vl_2|}^{\phi\phi} \, .
\end{align}

The local large-scale temperature gradient power can be expressed in harmonic space as
\begin{align}\label{eq:LSPowerDef}
    \lambda(\vL) = \int \frac{\dd^2\vl_3}{2\pi} h(\vl_3, \vL) \left(\vl_3 \cdot (\vl_3-\vL) \right) T(\vl_3) T(\vL-\vl_3) \, ,
\end{align}
where $h$ is a filter applied to the large-scale temperature fluctuations.
On large angular scales, lensing is only a small correction, and so we work to zeroth order in the lensing potential for the temperature fluctuations appearing in $\lambda$.

Our aim is to isolate the small scale lensing power by analyzing correlations between the $\varsigma$ and $\lambda$ fields.  The product of $\varsigma$ and $\lambda$ is
\begin{align}
    \varsigma(\vL) \lambda(\vL') =&
    \int \frac{\dd^2\vl_1}{2\pi} 
    g(\vl_1, \vL) \left(\vl_1 \cdot (\vl_1-\vL) \right) T(\vl_1) T(\vL-\vl_1)
    \nonumber \\
    &\times \int \frac{\dd^2\vl_3}{2\pi}
    h(\vl_3, \vLp) 
    \left(\vl_3 \cdot (\vl_3-\vLp) \right) 
    T(\vl_3) T(\vLp-\vl_3) \, .
\end{align}
The cross-spectrum of $\varsigma$ and $\lambda$ is given for $\vL\neq0$ by
\begin{align}
    \left \langle \varsigma(\vL) \lambda(\vL') \right \rangle=&
    \int \frac{\dd^2\vl_1}{2\pi} \int \frac{\dd^2\vl_3}{2\pi}
    g(\vl_1, \vL) \left(\vl_1 \cdot (\vl_1-\vL) \right)  h(\vl_3, \vLp) 
    \left(\vl_3 \cdot (\vl_3-\vLp) \right)
    \nonumber \\
    &\times 
    \left \langle T(\vl_1) T(\vL-\vl_1) T(\vl_3) T(\vLp-\vl_3) \right \rangle \, .
\end{align}
Choosing disjoint ranges of multipoles for the small-scale and large-scale temperature fluctuations means that the disconnected part of the temperature four-point function vanishes for $\vL\neq0$. The signal of interest is the one which is first order in the lensing power spectrum, whose dominant contribution comes from the terms of first order in the lensing potential in the gradient expansion of the small-scale temperature fluctuations
\begin{align}
    \left\langle \varsigma(\vL) \lambda(\vL') \right\rangle =&
    \int \frac{\dd^2\vl_1}{2\pi} 
    g(\vl_1, \vL) \left(\vl_1 \cdot (\vl_1-\vL) \right)  \int \frac{\dd^2\vl_2}{(2\pi)^2} 
    \left(\vl_2 \cdot (\vl_2-\vl_1) \right)
    \left( (\vL-\vl_2) \cdot (\vl_1-\vl_2) \right) 
    C_{|\vl_1-\vl_2|}^{\phi\phi}
    \nonumber \\
    &\times \int \frac{\dd^2\vl_3}{2\pi}
    h(\vl_3, \vL) 
    \left(\vl_3 \cdot (\vl_3-\vL) \right) 
    \left\langle  T(\vl_2) T(\vL-\vl_2) T(\vl_3) T(\vL-\vl_3) \right\rangle \delta(\vL+\vL') \nonumber \\
    =& \int \frac{\dd^2\vl_1}{(2\pi)^2} 
    g(\vl_1, \vL) \left(\vl_1 \cdot (\vl_1-\vL) \right)  \int \frac{\dd^2\vl_2}{(2\pi)^2} 
    \left(\vl_2 \cdot (\vl_2-\vl_1) \right)
    \left( (\vL-\vl_2) \cdot (\vl_1-\vl_2) \right) 
    C_{|\vl_1-\vl_2|}^{\phi\phi}
    \nonumber \\
    &\times 
    \left( 
    h(-\vl_2, -\vL) + h(\vl_2-\vL, -\vL) \right)
    \left(\vl_2 \cdot (\vl_2-\vL) \right)
    C_{\ell_2}^{TT} C_{|\vL-\vl_2|}^{TT} \delta(\vL+\vL') \nonumber \\
    =& C_\Lcheck^{\lambda\varsigma} \delta(\vL+\vL')\, .
\end{align}

We wish to use this cross-spectrum to obtain an unbiased estimate of the integrated small-scale lensing power.
We define a quantity
\begin{align}\label{eq:PsiL}
    %\Psi_\Lcheck \equiv A_\Lcheck C_\Lcheck^{GK} \, ,
    \Psi_\Lcheck \equiv A_\Lcheck \left\langle \varsigma(\vL{}) \lambda(-\vL) \right\rangle \, ,
\end{align}
with $A_\Lcheck$ defined such that $\Psi_\Lcheck$ is a weighted-average of the small-scale lensing power
\begin{align}\label{eq:ALDef}
    A_\Lcheck =& \left[ \int \frac{\dd^2\vl_1}{(2\pi)^2} 
    g(\vl_1, \vL) \left(\vl_1 \cdot (\vl_1-\vL) \right)  \int \frac{\dd^2\vl_2}{(2\pi)^2} 
    \left(\vl_2 \cdot (\vl_2-\vl_1) \right)
    \left( (\vL-\vl_2) \cdot (\vl_1-\vl_2) \right) \right. 
    \nonumber \\
    &\times \left.
    \left( 
    h(-\vl_2, -\vL) + h(\vl_2-\vL, -\vL) \right)
    \left(\vl_2 \cdot (\vl_2-\vL) \right)
    C_{\ell_2}^{TT} C_{|\vL-\vl_2|}^{TT} \right]^{-1} \, .
\end{align}

Let us briefly change to a full-sky notation which makes the calculation of the variance more transparent.  We wish to estimate the small-scale lensing power from the cross-spectrum of the $\varsigma$ and $\lambda$ fields
\begin{equation}
    A_\Lcheck \left\langle \varsigma_{\Lcheck \Mcheck} \lambda_{\Lcheck' -\Mcheck'} \right\rangle \equiv \Psi_\Lcheck \delta_{\Lcheck \Lcheck'} \delta_{\Mcheck \Mcheck'} \, .
\end{equation}
An estimator for $\Psi_\Lcheck$ can be constructed as
\begin{equation}
    \hat \Psi_\Lcheck \equiv A_\Lcheck \frac{1}{2\Lcheck+1} \sum_{\Mcheck} \varsigma_{\Lcheck \Mcheck} \lambda_{\Lcheck -\Mcheck} \, ,
\end{equation}
such that in an isotropic universe
\begin{equation}
    \left\langle \hat\Psi_\Lcheck \right\rangle = A_\Lcheck \frac{1}{2\Lcheck+1} \sum_{\Mcheck} \left\langle \varsigma_{\Lcheck \Mcheck} \lambda_{\Lcheck -\Mcheck} \right\rangle = \Psi_\Lcheck \, .
\end{equation}
The variance of this estimator can then be computed to be
\begin{align}
    \left\langle \left( \hat\Psi_\Lcheck - \Psi_\Lcheck \right)^2 \right\rangle 
    =& A_{\Lcheck}^2 \frac{1}{(2\Lcheck+1)^2} \sum_{\Mcheck \Mcheck'} \left\langle \varsigma_{\Lcheck \Mcheck} \lambda_{\Lcheck -\Mcheck} \varsigma_{\Lcheck \Mcheck'} \lambda_{\Lcheck -\Mcheck'} \right\rangle - \Psi_\Lcheck^2  \nonumber \\
    =& A_\Lcheck^2 \frac{1}{(2\Lcheck+1)^2}  \Bigg[ 
    \left (\sum_{\Mcheck} \left\langle \varsigma_{\Lcheck \Mcheck} \lambda_{\Lcheck -\Mcheck} \right \rangle \right)^2 
    + \sum_{\Mcheck} \left ( \left\langle \varsigma_{\Lcheck \Mcheck} \lambda_{\Lcheck -\Mcheck} \right \rangle \right)^2 \nonumber \\
    &+ \sum_{\Mcheck}  \left\langle \varsigma_{\Lcheck \Mcheck} \varsigma_{\Lcheck -\Mcheck} \right \rangle \left\langle \lambda_{\Lcheck \Mcheck} \lambda_{\Lcheck -\Mcheck} \right \rangle
    \Bigg] - \Psi_\Lcheck^2 \nonumber \\
    =& \frac{1}{2\Lcheck+1} \left[ \Psi_\Lcheck^2 + N_\Lcheck \right] \, ,
    \label{eq:Variance_fullsky}
\end{align}
where we have defined 
\begin{align}
    N_\Lcheck \equiv A_\Lcheck^2 \left\langle \varsigma_{\Lcheck \Mcheck} \varsigma_{\Lcheck -\Mcheck} \right \rangle \left\langle \lambda_{\Lcheck \Mcheck} \lambda_{\Lcheck -\Mcheck} \right \rangle \, .
    \label{eq:NL_def}
\end{align}

Next, we need to choose the filters $g$ and $h$ to minimize the variance of our small-scale lensing estimate.  Returning to the flat-sky approximation, the noise variance can be expressed as
\begin{align}
    N_\Lcheck =&
    A_\Lcheck^2 \int \frac{\dd^2\vl_1}{(2\pi)^2} 
    g(\vl_1, \vL)
    \left( g(-\vl_1, -\vL) + g(\vl_1-\vL, -\vL)  \right)
    \left(\vl_1 \cdot (\vl_1-\vL) \right)^2 C_{\ell_1}^{TT,\mathrm{obs}} C_{|\vL-\vl_1|}^{TT,\mathrm{obs}} \nonumber \\
    &\times    
    \int \frac{\dd^2\vl_2}{(2\pi)^2} 
    h(\vl_2, \vL)
    \left( h(-\vl_2, -\vL) + h(\vl_2-\vL, -\vL)  \right)
    \left(\vl_2 \cdot (\vl_2-\vL) \right)^2 C_{\ell_2}^{TT,\mathrm{obs}} C_{|\vL-\vl_2|}^{TT,\mathrm{obs}} \, .
    \label{eq:NL_flatsky}
\end{align}
Differentiating with respect to the choice of $g$ filter we find
\begin{align}
    %\frac{ \partial \mathrm{Var}(\Psi_\Lcheck) }{\partial g(\vl', \vL)}
    \frac{ \partial N_\Lcheck }{\partial g(\vl', \vL)}=&
    \frac{2 A_\Lcheck^2}{(2\pi)^2} \left( g(-\vl', -\vL) + g(\vl'-\vL, -\vL) \right)
    \left(\vl' \cdot (\vl'-\vL) \right)^2
    C_{\ell'}^{TT,\mathrm{obs}} C_{|\vL-\vl'|}^{TT,\mathrm{obs}} \nonumber \\
    & \quad \times    
    \int \frac{\dd^2\vl_2}{(2\pi)^2} 
    h(\vl_2, \vL)
    \left( h(-\vl_2, -\vL) + h(\vl_2-\vL, -\vL)  \right)
    \left(\vl_2 \cdot (\vl_2-\vL) \right)^2 C_{\ell_2}^{TT,\mathrm{obs}} C_{|\vL-\vl_2|}^{TT,\mathrm{obs}} \nonumber \\
    & - \frac{2 A_\Lcheck^3}{(2\pi^2)} 
    \left(\vl' \cdot (\vl'-\vL) \right)  \int \frac{\dd^2\vl_2}{(2\pi)^2} 
    \left(\vl_2 \cdot (\vl_2-\vl') \right)
    \left( (\vL-\vl_2) \cdot (\vl'-\vl_2) \right)
    \nonumber \\
    & \quad \times 
    \left( 
    h(-\vl_2, -\vL) + h(\vl_2-\vL, -\vL) \right)
    \left(\vl_2 \cdot (\vl_2-\vL) \right)
    C_{\ell_2}^{TT} C_{|\vL-\vl_2|}^{TT}
    \nonumber \\
    & \quad \times
    \int \frac{\dd^2\vl_1}{(2\pi)^2} 
    g(\vl_1, \vL)
    \left( g(-\vl_1, -\vL) + g(\vl_1-\vL, -\vL)  \right)
    \left(\vl_1 \cdot (\vl_1-\vL) \right)^2 C_{\ell_1}^{TT,\mathrm{obs}} C_{|\vL-\vl_1|}^{TT,\mathrm{obs}} \nonumber \\
    & \quad \times    
    \int \frac{\dd^2\vl_3}{(2\pi)^2} 
    h(\vl_3, \vL)
    \left( h(-\vl_3, -\vL) + h(\vl_3-\vL, -\vL)  \right)
    \left(\vl_3 \cdot (\vl_3-\vL) \right)^2 C_{\ell_3}^{TT,\mathrm{obs}} C_{|\vL-\vl_3|}^{TT,\mathrm{obs}} \, .
\end{align}
Setting this equal to zero and rearranging, we find
\begin{align}
    0 = &
    \left( g(-\vl', -\vL) + g(\vl'-\vL, -\vL) \right)
    \left(\vl' \cdot (\vl'-\vL) \right)^2
    C_{\ell'}^{TT,\mathrm{obs}} C_{|\vL-\vl'|}^{TT,\mathrm{obs}} \nonumber \\
    & \quad \times \int \frac{\dd^2\vl_1}{(2\pi)^2} 
    g(\vl_1, \vL) \left(\vl_1 \cdot (\vl_1-\vL) \right)  \int \frac{\dd^2\vl_2}{(2\pi)^2} 
    \left(\vl_2 \cdot (\vl_2-\vl_1) \right)
    \left( (\vL-\vl_2) \cdot (\vl_1-\vl_2) \right) 
    \nonumber \\
    & \quad \times 
    \left( 
    h(-\vl_2, -\vL) + h(\vl_2-\vL, -\vL) \right)
    \left(\vl_2 \cdot (\vl_2-\vL) \right)
    C_{\ell_2}^{TT} C_{|\vL-\vl_2|}^{TT} \nonumber \\
    & -  
    \left(\vl' \cdot (\vl'-\vL) \right)  \int \frac{\dd^2\vl_2}{(2\pi)^2} 
    \left(\vl_2 \cdot (\vl_2-\vl') \right)
    \left( (\vL-\vl_2) \cdot (\vl'-\vl_2) \right)
    \nonumber \\
    & \quad \times 
    \left( 
    h(-\vl_2, -\vL) + h(\vl_2-\vL, -\vL) \right)
    \left(\vl_2 \cdot (\vl_2-\vL) \right)
    C_{\ell_2}^{TT} C_{|\vL-\vl_2|}^{TT}
    \nonumber \\
    & \quad \times
    \int \frac{\dd^2\vl_1}{(2\pi)^2} 
    g(\vl_1, \vL)
    \left( g(-\vl_1, -\vL) + g(\vl_1-\vL, -\vL)  \right)
    \left(\vl_1 \cdot (\vl_1-\vL) \right)^2 C_{\ell_1}^{TT,\mathrm{obs}} C_{|\vL-\vl_1|}^{TT,\mathrm{obs}}
\end{align}
A similar procedure for the derivative with respect to the $h$ filter gives
\begin{align}
    0 = &
    \left( h(-\vl', -\vL) + h(\vl'-\vL, -\vL) \right)
    \left(\vl' \cdot (\vl'-\vL) \right)^2
    C_{\ell'}^{TT,\mathrm{obs}} C_{|\vL-\vl'|}^{TT,\mathrm{obs}} \nonumber \\
    & \quad \times \int \frac{\dd^2\vl_1}{(2\pi)^2} 
    g(\vl_1, \vL) \left(\vl_1 \cdot (\vl_1-\vL) \right)  \int \frac{\dd^2\vl_2}{(2\pi)^2} 
    \left(\vl_2 \cdot (\vl_2-\vl_1) \right)
    \left( (\vL-\vl_2) \cdot (\vl_1-\vl_2) \right) 
    \nonumber \\
    & \quad \times 
    \left( 
    h(-\vl_2, -\vL) + h(\vl_2-\vL, -\vL) \right)
    \left(\vl_2 \cdot (\vl_2-\vL) \right)
    C_{\ell_2}^{TT} C_{|\vL-\vl_2|}^{TT} \nonumber \\
    & -  
    \left(\vl' \cdot (\vl'-\vL) \right) C_{\ell'}^{TT} C_{|\vL-\vl'|}^{TT}  
    \nonumber \\
    & \quad \times \int \frac{\dd^2\vl_1}{(2\pi)^2} 
    g(\vl_1, \vL) \left(\vl_1 \cdot (\vl_1-\vL) \right)  
    \left(\vl' \cdot (\vl'-\vl_1) \right)
    \left( (\vL-\vl') \cdot (\vl_1-\vl') \right) \nonumber \\
    & \quad \times
    \int \frac{\dd^2\vl_2}{(2\pi)^2} 
    h(\vl_2, \vL)
    \left( h(-\vl_2, -\vL) + h(\vl_2-\vL, -\vL)  \right)
    \left(\vl_2 \cdot (\vl_2-\vL) \right)^2 C_{\ell_2}^{TT,\mathrm{obs}} C_{|\vL-\vl_2|}^{TT,\mathrm{obs}}
\end{align}

These equations are difficult to solve in general, but if we restrict attention to cases where $\varsigma$ includes only temperature fluctuations on scales much smaller than fluctuations appearing in $\lambda$ and also much smaller than scales defined by $\Lcheck$, then one can see that an approximate solution is provided by
\begin{align}
    g(\vl, \vL) &= W_\varsigma(\vl) \frac{1}{C_\ell^{TT,\mathrm{obs}}} W_\varsigma(\vL-\vl) \frac{1}{C_{|\vL-\vl|}^{TT,\mathrm{obs}}}   \, , 
    \label{eq:g_filter} \\
    h(\vl, \vL) &= W_\lambda(\vl) \frac{C_\ell^{TT}}{C_\ell^{TT,\mathrm{obs}}} W_\lambda(\vL-\vl) \frac{C_{|\vL-\vl|}^{TT}}{C_{|\vL-\vl|}^{TT,\mathrm{obs}}} \, ,
    \label{eq:h_filter}
\end{align}
where $W_\varsigma$ and $W_\lambda$ are window functions that restrict the temperature fluctuations to the appropriate scales,
\begin{align}
    W_\varsigma(\vl) =& 
    \begin{cases}
        1, & \ell_{1,\mathrm{min}} \leq |\vl| < \ell_{1,\mathrm{max}} \\
        0, & \text{else} \, ,
    \end{cases}  \label{eq:W_sigma} \\
    W_\lambda(\vl) =& 
    \begin{cases}
        1, & \ell_{2,\mathrm{min}} \leq |\vl| < \ell_{2,\mathrm{max}} \\
        0, & \text{else} \, .
    \end{cases} \label{eq:W_lambda}
\end{align}
Using this choice of filters gives
\begin{align}
    A_\Lcheck =& \left[ 2 \int \frac{\dd^2\vl_1}{(2\pi)^2} W_\varsigma(\vl_1) W_\varsigma(\vL-\vl_1)
    \left(\vl_1 \cdot (\vl_1-\vL) \right) \frac{1}{C_{\ell_1}^{TT,\mathrm{obs}}} \frac{1}{C_{|\vL-\vl_1|}^{TT,\mathrm{obs}}} 
    \phantom{\frac{\left( C_{|\vL-\vl_2|}^{TT} \right)^2} {C_{|\vL-\vl_2|}^{TT,\mathrm{obs}}}} \right. 
    \nonumber \\
    & \quad \times 
    \int \frac{\dd^2\vl_2}{(2\pi)^2} W_\lambda(\vl_2)W_\lambda(\vL-\vl_2)
    \left(\vl_2 \cdot (\vl_2-\vl_1) \right)
    \left( (\vL-\vl_2) \cdot (\vl_1-\vl_2) \right) \left(\vl_2 \cdot (\vl_2-\vL) \right) \nonumber \\
    & \qquad \times \left.
    \frac{\left( C_{\ell_2}^{TT} \right)^2}{C_{\ell_2}^{TT,\mathrm{obs}}} \frac{\left( C_{|\vL-\vl_2|}^{TT} \right)^2} {C_{|\vL-\vl_2|}^{TT,\mathrm{obs}}} \right]^{-1} \, , 
\end{align}
and the expected value of $\hat{\Psi}_\Lcheck$ is
\begin{align}\label{eq:PsiExpected}
    \left\langle \hat{\Psi}_\Lcheck \right\rangle =& 2 A_\Lcheck
    \int \frac{\dd^2\vl_1}{(2\pi)^2} W_\varsigma(\vl_1)W_\varsigma(\vL-\vl_1)
    \left(\vl_1 \cdot (\vl_1-\vL) \right) \frac{1}{C_{\ell_1}^{TT,\mathrm{obs}}} \frac{1}{C_{|\vL-\vl_1|}^{TT,\mathrm{obs}}}
    \nonumber \\
    & \quad \times 
    \int \frac{\dd^2\vl_2}{(2\pi)^2} W_\lambda(\vl_2)W_\lambda(\vL-\vl_2)
    \left(\vl_2 \cdot (\vl_2-\vl_1) \right)
    \left( (\vL-\vl_2) \cdot (\vl_1-\vl_2) \right) \left(\vl_2 \cdot (\vl_2-\vL) \right) \nonumber \\
    & \qquad \times
    \frac{\left( C_{\ell_2}^{TT} \right)^2}{C_{\ell_2}^{TT,\mathrm{obs}}} \frac{\left( C_{|\vL-\vl_2|}^{TT} \right)^2} {C_{|\vL-\vl_2|}^{TT,\mathrm{obs}}} C_{|\vl_1-\vl_2|}^{\phi\phi} \, , 
\end{align}
The noise variance of $\hat{\Psi}_\Lcheck$ is
\begin{align}
    %\mathrm{Var}(\Psi_\Lcheck) 
    N_\Lcheck =&
    4 A_\Lcheck^2 \int \frac{\dd^2\vl_1}{(2\pi)^2} W_\varsigma(\vl_1)W_\varsigma(\vL-\vl_1)
    \left(\vl_1 \cdot (\vl_1-\vL) \right)^2 \frac{1}{C_{\ell_1}^{TT,\mathrm{obs}}} \frac{1}{C_{|\vL-\vl_1|}^{TT,\mathrm{obs}}} \nonumber \\
    & \quad \times    
    \int \frac{\dd^2\vl_2}{(2\pi)^2} W_\lambda(\vl_2)W_\lambda(\vL-\vl_2)
    \left(\vl_2 \cdot (\vl_2-\vL) \right)^2 \frac{\left( C_{\ell_2}^{TT} \right)^2}{C_{\ell_2}^{TT,\mathrm{obs}}} \frac{\left( C_{|\vL-\vl_2|}^{TT} \right)^2}{C_{|\vL-\vl_2|}^{TT,\mathrm{obs}}} \nonumber \\
    \simeq & 4 A_\Lcheck \, ,
\end{align}
where in the last line, we used the same approximations as in deriving the $g$ and $h$ filters.

\twocolumngrid{}
\bibliographystyle{utphys}
\bibliography{patchbib} 

\providecommand{\href}[2]{#2}\begingroup\raggedright\begin{thebibliography}{10}

\bibitem{Lewis:2006fu}
A.~Lewis and A.~Challinor, ``{Weak gravitational lensing of the CMB},''
  \href{http://dx.doi.org/10.1016/j.physrep.2006.03.002}{{\em Phys. Rept.}
  {\bfseries 429} (2006) 1--65},
  \href{http://arxiv.org/abs/astro-ph/0601594}{{\ttfamily
  arXiv:astro-ph/0601594}}.

\bibitem{ACT:2020gnv}
{\bfseries ACT} Collaboration, S.~Aiola {\em et~al.}, ``{The Atacama Cosmology
  Telescope: DR4 Maps and Cosmological Parameters},''
  \href{http://dx.doi.org/10.1088/1475-7516/2020/12/047}{{\em JCAP} {\bfseries
  12} (2020) 047}, \href{http://arxiv.org/abs/2007.07288}{{\ttfamily
  arXiv:2007.07288 [astro-ph.CO]}}.

\bibitem{SPT:2017jdf}
{\bfseries SPT} Collaboration, J.~W. Henning {\em et~al.}, ``{Measurements of
  the Temperature and E-Mode Polarization of the CMB from 500 Square Degrees of
  SPTpol Data},'' \href{http://dx.doi.org/10.3847/1538-4357/aa9ff4}{{\em
  Astrophys. J.} {\bfseries 852} no.~2, (2018) 97},
  \href{http://arxiv.org/abs/1707.09353}{{\ttfamily arXiv:1707.09353
  [astro-ph.CO]}}.

\bibitem{Planck:2018nkj}
{\bfseries Planck} Collaboration, N.~Aghanim {\em et~al.}, ``{Planck 2018
  results. I. Overview and the cosmological legacy of Planck},''
  \href{http://dx.doi.org/10.1051/0004-6361/201833880}{{\em Astron. Astrophys.}
  {\bfseries 641} (2020) A1}, \href{http://arxiv.org/abs/1807.06205}{{\ttfamily
  arXiv:1807.06205 [astro-ph.CO]}}.

\bibitem{SPT:2014puc}
{\bfseries SPT} Collaboration, K.~T. Story {\em et~al.}, ``{A Measurement of
  the Cosmic Microwave Background Gravitational Lensing Potential from 100
  Square Degrees of SPTpol Data},''
  \href{http://dx.doi.org/10.1088/0004-637X/810/1/50}{{\em Astrophys. J.}
  {\bfseries 810} no.~1, (2015) 50},
  \href{http://arxiv.org/abs/1412.4760}{{\ttfamily arXiv:1412.4760
  [astro-ph.CO]}}.

\bibitem{Planck:2018lbu}
{\bfseries Planck} Collaboration, N.~Aghanim {\em et~al.}, ``{Planck 2018
  results. VIII. Gravitational lensing},''
  \href{http://dx.doi.org/10.1051/0004-6361/201833886}{{\em Astron. Astrophys.}
  {\bfseries 641} (2020) A8}, \href{http://arxiv.org/abs/1807.06210}{{\ttfamily
  arXiv:1807.06210 [astro-ph.CO]}}.

\bibitem{Darwish:2020fwf}
O.~Darwish {\em et~al.}, ``{The Atacama Cosmology Telescope: A CMB lensing mass
  map over 2100 square degrees of sky and its cross-correlation with BOSS-CMASS
  galaxies},'' \href{http://dx.doi.org/10.1093/mnras/staa3438}{{\em Mon. Not.
  Roy. Astron. Soc.} {\bfseries 500} no.~2, (2020) 2250--2263},
  \href{http://arxiv.org/abs/2004.01139}{{\ttfamily arXiv:2004.01139
  [astro-ph.CO]}}.

\bibitem{Hu:2001kj}
W.~Hu and T.~Okamoto, ``{Mass reconstruction with cmb polarization},''
  \href{http://dx.doi.org/10.1086/341110}{{\em Astrophys. J.} {\bfseries 574}
  (2002) 566--574}, \href{http://arxiv.org/abs/astro-ph/0111606}{{\ttfamily
  arXiv:astro-ph/0111606}}.

\bibitem{Okamoto:2003zw}
T.~Okamoto and W.~Hu, ``{CMB lensing reconstruction on the full sky},''
  \href{http://dx.doi.org/10.1103/PhysRevD.67.083002}{{\em Phys. Rev. D}
  {\bfseries 67} (2003) 083002},
  \href{http://arxiv.org/abs/astro-ph/0301031}{{\ttfamily
  arXiv:astro-ph/0301031}}.

\bibitem{Hirata:2003ka}
C.~M. Hirata and U.~Seljak, ``{Reconstruction of lensing from the cosmic
  microwave background polarization},''
  \href{http://dx.doi.org/10.1103/PhysRevD.68.083002}{{\em Phys. Rev. D}
  {\bfseries 68} (2003) 083002},
  \href{http://arxiv.org/abs/astro-ph/0306354}{{\ttfamily
  arXiv:astro-ph/0306354}}.

\bibitem{Smith:2010gu}
K.~M. Smith, D.~Hanson, M.~LoVerde, C.~M. Hirata, and O.~Zahn, ``{Delensing CMB
  Polarization with External Datasets},''
  \href{http://dx.doi.org/10.1088/1475-7516/2012/06/014}{{\em JCAP} {\bfseries
  06} (2012) 014}, \href{http://arxiv.org/abs/1010.0048}{{\ttfamily
  arXiv:1010.0048 [astro-ph.CO]}}.

\bibitem{Millea:2020iuw}
M.~Millea {\em et~al.}, ``{Optimal Cosmic Microwave Background Lensing
  Reconstruction and Parameter Estimation with SPTpol Data},''
  \href{http://dx.doi.org/10.3847/1538-4357/ac02bb}{{\em Astrophys. J.}
  {\bfseries 922} no.~2, (2021) 259},
  \href{http://arxiv.org/abs/2012.01709}{{\ttfamily arXiv:2012.01709
  [astro-ph.CO]}}.

\bibitem{Hirata:2002jy}
C.~M. Hirata and U.~Seljak, ``{Analyzing weak lensing of the cosmic microwave
  background using the likelihood function},''
  \href{http://dx.doi.org/10.1103/PhysRevD.67.043001}{{\em Phys. Rev. D}
  {\bfseries 67} (2003) 043001},
  \href{http://arxiv.org/abs/astro-ph/0209489}{{\ttfamily
  arXiv:astro-ph/0209489}}.

\bibitem{Horowitz:2017iql}
B.~Horowitz, S.~Ferraro, and B.~D. Sherwin, ``{Reconstructing Small Scale
  Lenses from the Cosmic Microwave Background Temperature Fluctuations},''
  \href{http://dx.doi.org/10.1093/mnras/stz566}{{\em Mon. Not. Roy. Astron.
  Soc.} {\bfseries 485} no.~3, (2019) 3919--3929},
  \href{http://arxiv.org/abs/1710.10236}{{\ttfamily arXiv:1710.10236
  [astro-ph.CO]}}.

\bibitem{Carron:2017mqf}
J.~Carron and A.~Lewis, ``{Maximum a posteriori CMB lensing reconstruction},''
  \href{http://dx.doi.org/10.1103/PhysRevD.96.063510}{{\em Phys. Rev. D}
  {\bfseries 96} no.~6, (2017) 063510},
  \href{http://arxiv.org/abs/1704.08230}{{\ttfamily arXiv:1704.08230
  [astro-ph.CO]}}.

\bibitem{Hadzhiyska:2019cle}
B.~Hadzhiyska, B.~D. Sherwin, M.~Madhavacheril, and S.~Ferraro, ``{Improving
  Small-Scale CMB Lensing Reconstruction},''
  \href{http://dx.doi.org/10.1103/PhysRevD.100.023547}{{\em Phys. Rev. D}
  {\bfseries 100} no.~2, (2019) 023547},
  \href{http://arxiv.org/abs/1905.04217}{{\ttfamily arXiv:1905.04217
  [astro-ph.CO]}}.

\bibitem{Dvorkin:2009}
C.~{Dvorkin} and K.~M. {Smith}, ``{Reconstructing patchy reionization from the
  cosmic microwave background},''
  \href{http://dx.doi.org/10.1103/PhysRevD.79.043003}{{\em \prd} {\bfseries 79}
  no.~4, (Feb., 2009) 043003}, \href{http://arxiv.org/abs/0812.1566}{{\ttfamily
  arXiv:0812.1566 [astro-ph]}}.

\bibitem{Hanson:2011}
D.~{Hanson}, A.~{Challinor}, G.~{Efstathiou}, and P.~{Bielewicz}, ``{CMB
  temperature lensing power reconstruction},''
  \href{http://dx.doi.org/10.1103/PhysRevD.83.043005}{{\em \prd} {\bfseries 83}
  no.~4, (Feb., 2011) 043005}, \href{http://arxiv.org/abs/1008.4403}{{\ttfamily
  arXiv:1008.4403 [astro-ph.CO]}}.

\bibitem{Kesden:2002ku}
M.~Kesden, A.~Cooray, and M.~Kamionkowski, ``{Separation of gravitational wave
  and cosmic shear contributions to cosmic microwave background
  polarization},'' \href{http://dx.doi.org/10.1103/PhysRevLett.89.011304}{{\em
  Phys. Rev. Lett.} {\bfseries 89} (2002) 011304},
  \href{http://arxiv.org/abs/astro-ph/0202434}{{\ttfamily
  arXiv:astro-ph/0202434}}.

\bibitem{Millea:2020cpw}
M.~Millea, E.~Anderes, and B.~D. Wandelt, ``{Sampling-based inference of the
  primordial CMB and gravitational lensing},''
  \href{http://dx.doi.org/10.1103/PhysRevD.102.123542}{{\em Phys. Rev. D}
  {\bfseries 102} no.~12, (2020) 123542},
  \href{http://arxiv.org/abs/2002.00965}{{\ttfamily arXiv:2002.00965
  [astro-ph.CO]}}.

\bibitem{Seljak:1999zn}
U.~Seljak and M.~Zaldarriaga, ``{Lensing induced cluster signatures in cosmic
  microwave background},'' \href{http://dx.doi.org/10.1086/309098}{{\em
  Astrophys. J.} {\bfseries 538} (2000) 57--64},
  \href{http://arxiv.org/abs/astro-ph/9907254}{{\ttfamily
  arXiv:astro-ph/9907254}}.

\bibitem{Knox:2002pe}
L.~Knox and Y.-S. Song, ``{A Limit on the detectability of the energy scale of
  inflation},'' \href{http://dx.doi.org/10.1103/PhysRevLett.89.011303}{{\em
  Phys. Rev. Lett.} {\bfseries 89} (2002) 011303},
  \href{http://arxiv.org/abs/astro-ph/0202286}{{\ttfamily
  arXiv:astro-ph/0202286}}.

\bibitem{Seljak:2003pn}
U.~Seljak and C.~M. Hirata, ``{Gravitational lensing as a contaminant of the
  gravity wave signal in CMB},''
  \href{http://dx.doi.org/10.1103/PhysRevD.69.043005}{{\em Phys. Rev. D}
  {\bfseries 69} (2004) 043005},
  \href{http://arxiv.org/abs/astro-ph/0310163}{{\ttfamily
  arXiv:astro-ph/0310163}}.

\bibitem{Green:2016cjr}
D.~Green, J.~Meyers, and A.~van Engelen, ``{CMB Delensing Beyond the B
  Modes},'' \href{http://dx.doi.org/10.1088/1475-7516/2017/12/005}{{\em JCAP}
  {\bfseries 12} (2017) 005}, \href{http://arxiv.org/abs/1609.08143}{{\ttfamily
  arXiv:1609.08143 [astro-ph.CO]}}.

\bibitem{Hotinli:2021umk}
S.~C. Hotinli, J.~Meyers, C.~Trendafilova, D.~Green, and A.~van Engelen, ``{The
  benefits of CMB delensing},''
  \href{http://dx.doi.org/10.1088/1475-7516/2022/04/020}{{\em JCAP} {\bfseries
  04} no.~04, (2022) 020}, \href{http://arxiv.org/abs/2111.15036}{{\ttfamily
  arXiv:2111.15036 [astro-ph.CO]}}.

\bibitem{Sherwin:2012mr}
B.~D. Sherwin {\em et~al.}, ``{The Atacama Cosmology Telescope:
  Cross-Correlation of CMB Lensing and Quasars},''
  \href{http://dx.doi.org/10.1103/PhysRevD.86.083006}{{\em Phys. Rev. D}
  {\bfseries 86} (2012) 083006},
  \href{http://arxiv.org/abs/1207.4543}{{\ttfamily arXiv:1207.4543
  [astro-ph.CO]}}.

\bibitem{HerschelATLAS:2014txv}
{\bfseries Herschel ATLAS} Collaboration, F.~Bianchini {\em et~al.},
  ``{Cross-correlation between the CMB lensing potential measured by Planck and
  high-z sub-mm galaxies detected by the Herschel-ATLAS survey},''
  \href{http://dx.doi.org/10.1088/0004-637X/802/1/64}{{\em Astrophys. J.}
  {\bfseries 802} no.~1, (2015) 64},
  \href{http://arxiv.org/abs/1410.4502}{{\ttfamily arXiv:1410.4502
  [astro-ph.CO]}}.

\bibitem{Liu:2015xfa}
J.~Liu and J.~C. Hill, ``{Cross-correlation of Planck CMB Lensing and CFHTLenS
  Galaxy Weak Lensing Maps},''
  \href{http://dx.doi.org/10.1103/PhysRevD.92.063517}{{\em Phys. Rev. D}
  {\bfseries 92} no.~6, (2015) 063517},
  \href{http://arxiv.org/abs/1504.05598}{{\ttfamily arXiv:1504.05598
  [astro-ph.CO]}}.

\bibitem{Schmittfull:2017ffw}
M.~Schmittfull and U.~Seljak, ``{Parameter constraints from cross-correlation
  of CMB lensing with galaxy clustering},''
  \href{http://dx.doi.org/10.1103/PhysRevD.97.123540}{{\em Phys. Rev. D}
  {\bfseries 97} no.~12, (2018) 123540},
  \href{http://arxiv.org/abs/1710.09465}{{\ttfamily arXiv:1710.09465
  [astro-ph.CO]}}.

\bibitem{Robertson:2020xom}
N.~C. Robertson {\em et~al.}, ``{Strong detection of the CMB lensing and galaxy
  weak lensing cross-correlation from ACT-DR4, Planck Legacy, and KiDS-1000},''
  \href{http://dx.doi.org/10.1051/0004-6361/202039975}{{\em Astron. Astrophys.}
  {\bfseries 649} (2021) A146},
  \href{http://arxiv.org/abs/2011.11613}{{\ttfamily arXiv:2011.11613
  [astro-ph.CO]}}.

\bibitem{Baxter:2022enq}
E.~J. Baxter {\em et~al.}, ``{Snowmass2021: Opportunities from Cross-survey
  Analyses of Static Probes},''
  \href{http://arxiv.org/abs/2203.06795}{{\ttfamily arXiv:2203.06795
  [hep-ex]}}.

\bibitem{DES:2022xxr}
{\bfseries DES, SPT} Collaboration, C.~Chang {\em et~al.}, ``{Joint analysis of
  Dark Energy Survey Year 3 data and CMB lensing from SPT and Planck. II.
  Cross-correlation measurements and cosmological constraints},''
  \href{http://dx.doi.org/10.1103/PhysRevD.107.023530}{{\em Phys. Rev. D}
  {\bfseries 107} no.~2, (2023) 023530},
  \href{http://arxiv.org/abs/2203.12440}{{\ttfamily arXiv:2203.12440
  [astro-ph.CO]}}.

\bibitem{Lin:2020sbb}
X.~Lin, Z.~Cai, Y.~Li, A.~Krolewski, and S.~Ferraro, ``{Constraining the Halo
  Mass of Damped Ly$\alpha$ Absorption Systems (DLAs) at z = 2\textendash{}3.5
  Using the Quasar-CMB Lensing Cross-correlation},''
  \href{http://dx.doi.org/10.3847/1538-4357/abc620}{{\em Astrophys. J.}
  {\bfseries 905} no.~2, (2020) 176},
  \href{http://arxiv.org/abs/2011.01234}{{\ttfamily arXiv:2011.01234
  [astro-ph.CO]}}.

\bibitem{Piccirilli:2022myi}
G.~Piccirilli, M.~Migliaccio, E.~Branchini, and A.~Dolfi, ``{A
  cross-correlation analysis of CMB lensing and radio galaxy maps},''
  \href{http://arxiv.org/abs/2208.07774}{{\ttfamily arXiv:2208.07774
  [astro-ph.CO]}}.

\bibitem{Kaplinghat:2003bh}
M.~Kaplinghat, L.~Knox, and Y.-S. Song, ``{Determining neutrino mass from the
  CMB alone},'' \href{http://dx.doi.org/10.1103/PhysRevLett.91.241301}{{\em
  Phys. Rev. Lett.} {\bfseries 91} (2003) 241301},
  \href{http://arxiv.org/abs/astro-ph/0303344}{{\ttfamily
  arXiv:astro-ph/0303344}}.

\bibitem{Lesgourgues:2012uu}
J.~Lesgourgues and S.~Pastor, ``{Neutrino mass from Cosmology},''
  \href{http://dx.doi.org/10.1155/2012/608515}{{\em Adv. High Energy Phys.}
  {\bfseries 2012} (2012) 608515},
  \href{http://arxiv.org/abs/1212.6154}{{\ttfamily arXiv:1212.6154 [hep-ph]}}.

\bibitem{Green:2021xzn}
D.~Green and J.~Meyers, ``{Cosmological Implications of a Neutrino Mass
  Detection},'' \href{http://arxiv.org/abs/2111.01096}{{\ttfamily
  arXiv:2111.01096 [astro-ph.CO]}}.

\bibitem{Abazajian:2022ofy}
K.~N. Abazajian {\em et~al.}, ``{Synergy between cosmological and laboratory
  searches in neutrino physics: a white paper},''
  \href{http://arxiv.org/abs/2203.07377}{{\ttfamily arXiv:2203.07377
  [hep-ph]}}.

\bibitem{Tulin:2017ara}
S.~Tulin and H.-B. Yu, ``{Dark Matter Self-interactions and Small Scale
  Structure},'' \href{http://dx.doi.org/10.1016/j.physrep.2017.11.004}{{\em
  Phys. Rept.} {\bfseries 730} (2018) 1--57},
  \href{http://arxiv.org/abs/1705.02358}{{\ttfamily arXiv:1705.02358
  [hep-ph]}}.

\bibitem{Gluscevic:2019yal}
V.~Gluscevic {\em et~al.}, ``{Cosmological Probes of Dark Matter Interactions:
  The Next Decade},'' {\em Bull. Am. Astron. Soc.} {\bfseries 51} no.~3, (2019)
  134, \href{http://arxiv.org/abs/1903.05140}{{\ttfamily arXiv:1903.05140
  [astro-ph.CO]}}.

\bibitem{Buen-Abad:2021mvc}
M.~A. Buen-Abad, R.~Essig, D.~McKeen, and Y.-M. Zhong, ``{Cosmological
  constraints on dark matter interactions with ordinary matter},''
  \href{http://dx.doi.org/10.1016/j.physrep.2022.02.006}{{\em Phys. Rept.}
  {\bfseries 961} (2022) 1--35},
  \href{http://arxiv.org/abs/2107.12377}{{\ttfamily arXiv:2107.12377
  [astro-ph.CO]}}.

\bibitem{Hui:2016ltb}
L.~Hui, J.~P. Ostriker, S.~Tremaine, and E.~Witten, ``{Ultralight scalars as
  cosmological dark matter},''
  \href{http://dx.doi.org/10.1103/PhysRevD.95.043541}{{\em Phys. Rev. D}
  {\bfseries 95} no.~4, (2017) 043541},
  \href{http://arxiv.org/abs/1610.08297}{{\ttfamily arXiv:1610.08297
  [astro-ph.CO]}}.

\bibitem{Ferreira:2020fam}
E.~G.~M. Ferreira, ``{Ultra-light dark matter},''
  \href{http://dx.doi.org/10.1007/s00159-021-00135-6}{{\em Astron. Astrophys.
  Rev.} {\bfseries 29} no.~1, (2021) 7},
  \href{http://arxiv.org/abs/2005.03254}{{\ttfamily arXiv:2005.03254
  [astro-ph.CO]}}.

\bibitem{Drewes:2016upu}
M.~Drewes {\em et~al.}, ``{A White Paper on keV Sterile Neutrino Dark
  Matter},'' \href{http://dx.doi.org/10.1088/1475-7516/2017/01/025}{{\em JCAP}
  {\bfseries 01} (2017) 025}, \href{http://arxiv.org/abs/1602.04816}{{\ttfamily
  arXiv:1602.04816 [hep-ph]}}.

\bibitem{Chisari:2019tus}
N.~E. Chisari {\em et~al.}, ``{Modelling baryonic feedback for survey
  cosmology},'' \href{http://dx.doi.org/10.21105/astro.1905.06082}{{\em Open J.
  Astrophys.} {\bfseries 2} no.~1, (2019) 4},
  \href{http://arxiv.org/abs/1905.06082}{{\ttfamily arXiv:1905.06082
  [astro-ph.CO]}}.

\bibitem{CMB-S4:2016ple}
{\bfseries CMB-S4} Collaboration, K.~N. Abazajian {\em et~al.}, ``{CMB-S4
  Science Book, First Edition},''
  \href{http://arxiv.org/abs/1610.02743}{{\ttfamily arXiv:1610.02743
  [astro-ph.CO]}}.

\bibitem{SimonsObservatory:2018koc}
{\bfseries Simons Observatory} Collaboration, P.~Ade {\em et~al.}, ``{The
  Simons Observatory: Science goals and forecasts},''
  \href{http://dx.doi.org/10.1088/1475-7516/2019/02/056}{{\em JCAP} {\bfseries
  02} (2019) 056}, \href{http://arxiv.org/abs/1808.07445}{{\ttfamily
  arXiv:1808.07445 [astro-ph.CO]}}.

\bibitem{NASAPICO:2019thw}
{\bfseries NASA PICO} Collaboration, S.~Hanany {\em et~al.}, ``{PICO: Probe of
  Inflation and Cosmic Origins},''
  \href{http://arxiv.org/abs/1902.10541}{{\ttfamily arXiv:1902.10541
  [astro-ph.IM]}}.

\bibitem{Abazajian:2019eic}
K.~Abazajian {\em et~al.}, ``{CMB-S4 Science Case, Reference Design, and
  Project Plan},'' \href{http://arxiv.org/abs/1907.04473}{{\ttfamily
  arXiv:1907.04473 [astro-ph.IM]}}.

\bibitem{Sehgal:2019ewc}
N.~Sehgal {\em et~al.}, ``{CMB-HD: An Ultra-Deep, High-Resolution
  Millimeter-Wave Survey Over Half the Sky},''
  \href{http://arxiv.org/abs/1906.10134}{{\ttfamily arXiv:1906.10134
  [astro-ph.CO]}}.

\bibitem{Hu:2007bt}
W.~Hu, S.~DeDeo, and C.~Vale, ``{Cluster Mass Estimators from CMB Temperature
  and Polarization Lensing},''
  \href{http://dx.doi.org/10.1088/1367-2630/9/12/441}{{\em New J. Phys.}
  {\bfseries 9} (2007) 441},
  \href{http://arxiv.org/abs/astro-ph/0701276}{{\ttfamily
  arXiv:astro-ph/0701276}}.

\bibitem{Smith:2016lnt}
K.~M. Smith and S.~Ferraro, ``{Detecting Patchy Reionization in the Cosmic
  Microwave Background},''
  \href{http://dx.doi.org/10.1103/PhysRevLett.119.021301}{{\em Phys. Rev.
  Lett.} {\bfseries 119} no.~2, (2017) 021301},
  \href{http://arxiv.org/abs/1607.01769}{{\ttfamily arXiv:1607.01769
  [astro-ph.CO]}}.

\bibitem{Kesden}
M.~{Kesden}, A.~{Cooray}, and M.~{Kamionkowski}, ``{Lensing reconstruction with
  CMB temperature and polarization},''
  \href{http://dx.doi.org/10.1103/PhysRevD.67.123507}{{\em \prd} {\bfseries 67}
  no.~12, (June, 2003) 123507},
  \href{http://arxiv.org/abs/astro-ph/0302536}{{\ttfamily
  arXiv:astro-ph/0302536 [astro-ph]}}.

\bibitem{Lewis:1999bs}
A.~Lewis, A.~Challinor, and A.~Lasenby, ``{Efficient computation of CMB
  anisotropies in closed FRW models},''
  \href{http://dx.doi.org/10.1086/309179}{{\em \apj} {\bfseries 538} (2000)
  473--476},
\href{http://arxiv.org/abs/astro-ph/9911177}{{\ttfamily arXiv:astro-ph/9911177
  [astro-ph]}}.
%%CITATION = ASTRO-PH/9911177;%%.

\bibitem{Howlett:2012mh}
C.~Howlett, A.~Lewis, A.~Hall, and A.~Challinor, ``{CMB power spectrum
  parameter degeneracies in the era of precision cosmology},''
  \href{http://dx.doi.org/10.1088/1475-7516/2012/04/027}{{\em \jcap} {\bfseries
  1204} (2012) 027},
\href{http://arxiv.org/abs/1201.3654}{{\ttfamily arXiv:1201.3654
  [astro-ph.CO]}}.
%%CITATION = ARXIV:1201.3654;%%.

\bibitem{McCarthy:2022}
F.~{McCarthy}, J.~C. {Hill}, and M.~S. {Madhavacheril}, ``{Baryonic feedback
  biases on fundamental physics from lensed CMB power spectra},''
  \href{http://dx.doi.org/10.1103/PhysRevD.105.023517}{{\em \prd} {\bfseries
  105} no.~2, (Jan., 2022) 023517},
  \href{http://arxiv.org/abs/2103.05582}{{\ttfamily arXiv:2103.05582
  [astro-ph.CO]}}.

\bibitem{Raghunathan:2018}
S.~{Raghunathan}, F.~{Bianchini}, and C.~L. {Reichardt}, ``{Imprints of
  gravitational lensing in the Planck cosmic microwave background data at the
  location of W I S E {\texttimes}SCOS galaxies},''
  \href{http://dx.doi.org/10.1103/PhysRevD.98.043506}{{\em \prd} {\bfseries 98}
  no.~4, (Aug., 2018) 043506},
  \href{http://arxiv.org/abs/1710.09770}{{\ttfamily arXiv:1710.09770
  [astro-ph.CO]}}.

\bibitem{DES:2017fyz}
{\bfseries DES, SPT} Collaboration, E.~J. Baxter {\em et~al.}, ``{A measurement
  of CMB cluster lensing with SPT and DES year 1 data},''
  \href{http://dx.doi.org/10.1093/mnras/sty305}{{\em Mon. Not. Roy. Astron.
  Soc.} {\bfseries 476} no.~2, (2018) 2674--2688},
  \href{http://arxiv.org/abs/1708.01360}{{\ttfamily arXiv:1708.01360
  [astro-ph.CO]}}.

\bibitem{ACT:2020izl}
{\bfseries ACT} Collaboration, M.~S. Madhavacheril {\em et~al.}, ``{The Atacama
  Cosmology Telescope: Weighing Distant Clusters with the Most Ancient
  Light},'' \href{http://dx.doi.org/10.3847/2041-8213/abbccb}{{\em Astrophys.
  J. Lett.} {\bfseries 903} no.~1, (2020) L13},
  \href{http://arxiv.org/abs/2009.07772}{{\ttfamily arXiv:2009.07772
  [astro-ph.CO]}}.

\bibitem{Han:2022}
D.~{Han} and N.~{Sehgal}, ``{Mitigating foreground bias to the CMB lensing
  power spectrum for a CMB-HD survey},''
  \href{http://dx.doi.org/10.1103/PhysRevD.105.083516}{{\em \prd} {\bfseries
  105} no.~8, (Apr., 2022) 083516},
  \href{http://arxiv.org/abs/2112.02109}{{\ttfamily arXiv:2112.02109
  [astro-ph.CO]}}.

\bibitem{vanEngelen:2013rla}
A.~van Engelen, S.~Bhattacharya, N.~Sehgal, G.~P. Holder, O.~Zahn, and
  D.~Nagai, ``{CMB Lensing Power Spectrum Biases from Galaxies and Clusters
  using High-angular Resolution Temperature Maps},''
  \href{http://dx.doi.org/10.1088/0004-637X/786/1/13}{{\em Astrophys. J.}
  {\bfseries 786} (2014) 13}, \href{http://arxiv.org/abs/1310.7023}{{\ttfamily
  arXiv:1310.7023 [astro-ph.CO]}}.

\bibitem{Osborne:2013nna}
S.~J. Osborne, D.~Hanson, and O.~Dor\'e, ``{Extragalactic Foreground
  Contamination in Temperature-based CMB Lens Reconstruction},''
  \href{http://dx.doi.org/10.1088/1475-7516/2014/03/024}{{\em JCAP} {\bfseries
  03} (2014) 024}, \href{http://arxiv.org/abs/1310.7547}{{\ttfamily
  arXiv:1310.7547 [astro-ph.CO]}}.

\bibitem{Madhavacheril:2018}
M.~S. {Madhavacheril} and J.~C. {Hill}, ``{Mitigating foreground biases in CMB
  lensing reconstruction using cleaned gradi ents},''
  \href{http://dx.doi.org/10.1103/PhysRevD.98.023534}{{\em \prd} {\bfseries 98}
  no.~2, (July, 2018) 023534},
  \href{http://arxiv.org/abs/1802.08230}{{\ttfamily arXiv:1802.08230
  [astro-ph.CO]}}.

\bibitem{Beck:2020}
D.~{Beck}, J.~{Errard}, and R.~{Stompor}, ``{Impact of polarized galactic
  foreground emission on CMB lensing reconstruction and delensing of
  B-modes},'' \href{http://dx.doi.org/10.1088/1475-7516/2020/06/030}{{\em
  \jcap} {\bfseries 2020} no.~6, (June, 2020) 030},
  \href{http://arxiv.org/abs/2001.02641}{{\ttfamily arXiv:2001.02641
  [astro-ph.CO]}}.

\bibitem{Abylkairov:2021}
Y.~S. {Abylkairov}, O.~{Darwish}, J.~C. {Hill}, and B.~D. {Sherwin},
  ``{Partially constrained internal linear combination: A method for low-noise
  CMB foreground mitigation},''
  \href{http://dx.doi.org/10.1103/PhysRevD.103.103510}{{\em \prd} {\bfseries
  103} no.~10, (May, 2021) 103510},
  \href{http://arxiv.org/abs/2012.04032}{{\ttfamily arXiv:2012.04032
  [astro-ph.CO]}}.

\bibitem{Darwish:2021}
O.~{Darwish}, B.~D. {Sherwin}, N.~{Sailer}, E.~{Schaan}, and S.~{Ferraro},
  ``{Optimizing foreground mitigation for CMB lensing with combined
  multifrequency and geometric methods},''
  \href{http://dx.doi.org/10.48550/arXiv.2111.00462}{{\em arXiv e-prints}
  (Oct., 2021) arXiv:2111.00462},
  \href{http://arxiv.org/abs/2111.00462}{{\ttfamily arXiv:2111.00462
  [astro-ph.CO]}}.

\bibitem{Mirmelstein:2020pfk}
M.~Mirmelstein, G.~Fabbian, A.~Lewis, and J.~Peloton, ``{Instrumental
  systematics biases in CMB lensing reconstruction: A simulation-based
  assessment},'' \href{http://dx.doi.org/10.1103/PhysRevD.103.123540}{{\em
  Phys. Rev. D} {\bfseries 103} no.~12, (2021) 123540},
  \href{http://arxiv.org/abs/2011.13910}{{\ttfamily arXiv:2011.13910
  [astro-ph.CO]}}.

\end{thebibliography}\endgroup

\end{document}